\documentclass[preprintnumbers,superscriptaddress,onecolumn,jcp,12pt]{revtex4-1}
\usepackage{graphicx}
\usepackage{amsmath}
\usepackage{mhchem}
\usepackage{longtable}
\usepackage{array}
\usepackage[left=0.5cm,right=0.5cm, top=2cm, bottom=2cm]{geometry}

\begin{document}

\title{Long-range corrected hybrid meta-generalized-gradient approximations with dispersion corrections}

\author{You-Sheng Lin}
\affiliation{Department of Physics, National Taiwan University, Taipei 10617, Taiwan}

\author{Chen-Wei Tsai}
\affiliation{Department of Physics, National Taiwan University, Taipei 10617, Taiwan}

\author{Guan-De Li}
\affiliation{Department of Physics, National Taiwan University, Taipei 10617, Taiwan}

\author{Jeng-Da Chai}
\email[Author to whom correspondence should be addressed. Electronic mail: ]{jdchai@phys.ntu.edu.tw.} 
\affiliation{Department of Physics, National Taiwan University, Taipei 10617, Taiwan}
\affiliation{Center for Theoretical Sciences and Center for Quantum Science and Engineering, National Taiwan University, Taipei 10617, Taiwan}

\date{\today{}}

\begin{abstract}

We propose a long-range corrected hybrid meta-GGA functional, based on a global hybrid meta-GGA functional, M05 [Y. Zhao, N. E. Schultz, and D. G. Truhlar, J. Chem. Phys. {\bf123}, 161103 (2005)], and 
empirical atom-atom dispersion corrections. Our resulting functional, $\omega\text{M05-D}$, is shown to be accurate for a very wide range of applications, such as thermochemistry, kinetics, noncovalent interactions, 
equilibrium geometries, frontier orbital energies, fundamental gaps, and excitation energies. In addition, we present three new databases, IP131 (131 ionization potentials), EA115 (115 electron affinities), and FG115 (115 fundamental gaps), consisting of experimental molecular geometries and accurate reference values, which will be useful in the assessment of the accuracy of density functional approximations.

\end{abstract}

\maketitle

\section{Introduction}\label{I}

Because of its satisfactory accuracy and modest cost in many applications, Kohn-Sham density functional theory (KS-DFT) \cite{Hohenberg64,Kohn65} has become one of the most popular electronic structure methods for 
large ground-state systems \cite{Parr89,Dreizler90,Engel11,Kohn96}. Its extension for treating excited-state systems, time-dependent density functional theory (TDDFT) \cite{Casida95,Gross96}, has also been widely used.

The crucial ingredient of KS-DFT, the exact exchange-correlation (XC) energy functional $E_{xc}[\rho]$, however, remains unknown and needs to be approximated. Functionals based on the local density approximation 
(LDA), modeling the XC energy density locally with that of a uniform electron gas (UEG), have been quite successful for nearly-free electron systems \cite{Parr89,Dreizler90}, though still insufficiently accurate for 
most quantum chemical applications. Functionals based on the generalized gradient approximations (GGAs), additionally incorporating the gradient of local density into the LDA, have achieved reasonable accuracy in many applications. As an extension of the GGA (for rather restricted set of density variables), meta-GGA (MGGA) offers itself quite naturally. Functionals depending directly on the Laplacian of the density have not been pursued intensively, because of the difficulty of numerical evaluation. MGGAs, which adopt the kinetic energy density as a substitute for the Laplacian, have shown evidences of superiority over GGAs. \cite{VS98,Tao03,M06L}

However, the LDA, GGAs and MGGAs (commonly denoted as DFAs for density functional approximations) are based on the localized model XC holes, while the exact XC hole should be fully nonlocal. Currently, perhaps the most 
successful approaches to taking into account the nonlocality of XC hole are provided by hybrid DFT methods, incorporating a fraction of the exact Hartree-Fock (HF) exchange into the DFAs. Hybrid density functionals 
have achieved remarkable accuracy and have expanded the usefulness of DFT for many applications. Noticeably, global hybrid MGGA functionals \cite{Becke96,Boese02,Staroverov03,Zhao04,Hill06,Boese04,Zhao05,Zhao06,Zhao_TCA08}, where the XC energy density depends on the local density, the gradient of local density, a fraction of exact exchange, 
as well as the exact KS kinetic energy density (a function of the occupied KS orbitals) \cite{Becke83,Becke98,Becke00,Schmider00}, have been shown to potentially perform better than 
global hybrid GGA functionals \cite{Zhao04,Hill06,Boese04,Zhao05,Zhao06,Zhao_TCA08,Zhao_ACR08,Zhao08}, due to the additional ingredient of kinetic energy density in global hybrid MGGA functionals.

In global hybrid functionals, a small fraction of the exact HF exchange is added to a semilocal density functional. In certain situations, especially in the asymptotic regions of molecular systems, a large fraction (even 100\%) 
of HF exchange is needed. Aiming to remedy this, long-range corrected (LC) hybrid DFT schemes have been actively developed \cite{Iikura01,Tawada04,Gerber05,Gerber07,Vydrov06,Vydrov06_2,Song07,Cohen07,Chai08,Chai08_2,Chai09}. LC hybrids retain the full HF exchange only for the long-range electron-electron interactions, and 
thereby resolve a significant part of the self-interaction problems associated with global hybrid functionals.

On the other hand, the development of accurate short-range (SR) exchange density functionals $E_x^{\text{SR}}[\rho]$, plays an important role in the progress of LC-DFT. In the first LC scheme, an ansatz for the conversion of any $E_x$ to 
$E_x^{\text{SR}}$ was proposed by Iikura {\it et al.} \cite{Iikura01}, and has become widely used. However, their resulting LC hybrid GGA functionals do not outperform the corresponding global hybrid GGA functionals for 
thermochemistry. In 2006, Vydrov {\it et al.} proposed a different LC scheme \cite{Vydrov06}, based on integrating a GGA model exchange hole. Their resulting LC-$\omega$PBE functional has shown improved performance for 
thermochemistry and barrier heights, and is comparable to global hybrid GGA functionals such as B3LYP \cite{Becke93,Stephens94}. However, further improvements following this direction require the development of 
more accurate model exchange holes, which is a quite challenging task.

Another approach to more accurate LC hybrid functionals was proposed by Chai and Head-Gordon \cite{Chai08}. First, augmenting the SR local spin density exchange energy density by a flexible enhancement factor 
(of the Becke's 1997 form \cite{B97}) and fully reoptimizing the LC functional on a diverse training set, yields the $\omega$B97 functional. Second, including an adjustable fraction of SR HF exchange in the $\omega$B97 
functional with the similar reoptimization procedure, leads to the $\omega$B97X functional. $\omega$B97 and $\omega$B97X have been shown to be accurate across a diverse set of test data, containing thermochemistry, 
kinetics, and noncovalent interactions \cite{Chai08}.

However, problems associated with the lack of nonlocality of the DFA correlation hole, such as the lack of dispersion interactions (the missing of van der Waals forces), are not resolved by the LC hybrid schemes. The correlation 
functionals in typical LC hybrids are treated semilocally, which cannot capture the long-range (LR) correlation effects \cite{Dobson01,Kristyan94}. To remedy this, the DFT-D scheme was applied \cite{Wu01,*Wu02,*Zimmerli04,*Grimme04,*Grimme06,*Antony06,*Jurecka06,*Goursot07,*Grimme07,*Cerny07,*Morgado07,*Kabelac07,*Cerny_PCCP07} to extend the $\omega$B97X functional with damped atom-atom dispersion corrections, denoted as $\omega$B97X-D \cite{Chai08_2}. Consequently, $\omega$B97X-D can obtain dispersive effects with essentially zero additional computational cost relative to $\omega$B97X. As an alternative approach, $\omega$B97X has also been combined with the double-hybrid methods \cite{Grimme_JCP06,Schwabe07,Tarnopolsky08,Benighaus08,Zhang09}, which mix both the HF exchange and nonlocal orbital correlation energy 
from the second-order perturbation energy expression in wave function theory. The resulting $\omega$B97X-2 functional \cite{Chai09} has yielded very high accuracy for thermochemistry, kinetics, and noncovalent interactions, 
though its fifth-order scaling with respect to system size may limit its applicability to larger systems. 

As the $\omega$B97 series are LC hybrid GGAs, it seems a natural step to develop LC hybrid MGGAs and to assess their performance. In this work, we propose a new LC hybrid MGGA-D functional, 
denoted as $\omega\text{M05-D}$, which is shown to be accurate for a wide range of applications, when compared with the two closely related functionals: a global hybrid MGGA functional (M05-2X) \cite{Zhao06} and 
a LC hybrid GGA-D functional ($\omega$B97X-D) \cite{Chai08_2}. The rest of this paper is organized as follows. In Sec. II, we briefly describe the relevant schemes developed in the LC hybrid approach. In Sec. III, we propose 
a new SR exchange functional, which serves as suitable basis functionals for systematically generating accurate LC hybrid MGGA functionals. The performance of the $\omega\text{M05-D}$ functional is compared with other 
functionals in Sec. IV (on the training set), and in Sec. V (on some test sets). In Sec. VI, we give our conclusions. 

\section{Rationales of LC Hybrid Schemes}

For the LC hybrid schemes, one first defines the long-range and short-range operators to partition the Coulomb operator. The most popular type of splitting operator used is the standard error function (erf),
\begin{equation}\label{splitting}
\frac{1}{r_{12}}=\frac{\text{erf}(\omega r_{12})}{r_{12}}+\frac{\text{erfc}(\omega r_{12})}{r_{12}},
\end{equation}
where $r_{12}\equiv\left|{\bf r}_{12}\right|=\left|{\bf r}_{1}-{\bf r}_{2}\right|$ (atomic units are used throughout this paper). On the right hand side of Eq. (\ref{splitting}), the first term is long-ranged, while the second term is short-ranged. 
The parameter $\omega$ defines the range of these operators. 

In this work, we employ the erf/erfc partition, and use the following expression (as suggested in the recent LC hybrid schemes \cite{Chai08,Chai08_2,M11}) 
for the LC hybrid functionals ($c_x$ is a fractional number to be determined):
\begin{equation}
E_{xc}^{\text{LC-DFA}} = E_{x}^{\text{LR-HF}} + c_{x} E_{x}^{\text{SR-HF}} + (1-c_{x}) E_{x}^{\text{SR-DFA}}+E_{c}^{\text{DFA}},
\label{eq:lcdft1}
\end{equation}
where $E_x^{\text{LR-HF}}$, the LR HF exchange, is computed by the occupied KS orbitals $\psi_{i\sigma}({\bf r})$ with the LR operator,
\begin{equation}
E_x^{\text{LR-HF}}=-\frac{1}{2}\sum_\sigma\sum_{i,j}^{\text{occ.}}\iint\psi^*_{i\sigma}({\bf r}_1)\psi^*_{j\sigma}({\bf r}_2)\frac{\mbox{erf}(\omega r_{12})}{r_{12}}\psi_{j\sigma}({\bf r}_1)\psi_{i\sigma}({\bf r}_2)d{\bf r}_1d{\bf r}_2,
\end{equation}
$E_x^{\text{SR-HF}}$, the SR HF exchange, is computed similarly to the above but with the SR operator,
\begin{equation}
E_x^{\text{SR-HF}}=-\frac{1}{2}\sum_\sigma\sum_{i,j}^{\text{occ.}}\iint\psi^*_{i\sigma}({\bf r}_1)\psi^*_{j\sigma}({\bf r}_2)\frac{\mbox{erfc}(\omega r_{12})}{r_{12}}\psi_{j\sigma}({\bf r}_1)\psi_{i\sigma}({\bf r}_2)d{\bf r}_1d{\bf r}_2,
\end{equation}
$E_{x}^{\text{SR-DFA}}$ is the SR exchange approximated by DFAs , and $E_{c}^{\text{DFA}}$ is the correlation functional the same as that of the full Coulomb interaction.  

In view of the $E_{xc}^{\text{LC-DFA}}$ in Eq.\ (\ref{eq:lcdft1}), as $E_x^{\text{LR-HF}}$ and $E_x^{\text{SR-HF}}$ are well defined, and accurate approximations for $E_{c}^{\text{DFA}}$ are widely available, the accuracy of 
$E_{x}^{\text{SR-DFA}}$ is thus closely related to the accuracy of a LC hybrid functional \cite{Chai08,Chai08_2}. The analytical form of the SR LDA (the simplest SR DFA) exchange functional $E_x^{\text{SR-LDA}}$ can be 
obtained by the integration of the square of the LDA density matrix with the SR operator \cite{Gill96},
\begin{equation}
E_x^{\text{SR-LDA}}=\sum_\sigma\int e_{x\sigma}^\text{SR-LDA}(\rho_\sigma)d{\bf r}.
\end{equation}
Here, $e_{x\sigma}^\text{SR-LDA}(\rho_\sigma)$ is the SR LDA exchange energy density for $\sigma$-spin,
\begin{equation}\label{eSR_LDA}
e_{x\sigma}^\text{SR-LDA}(\rho_\sigma)=e_{x\sigma}^\text{LDA}F(a_\sigma),
\end{equation}
where 
\begin{equation}
e_{x\sigma}^\text{LDA}(\rho_\sigma) = -\frac{3}{2}\left(\frac{3}{4\pi}\right)^{1/3}\rho_\sigma^{4/3}({\bf r})
\end{equation}
is the LDA exchange energy density for $\sigma$-spin, 
$k_{F\sigma}\equiv(6\pi^2\rho_\sigma({\bf r}))^{1/3}$ is the local Fermi wave vector, and $a_\sigma\equiv\omega/(2k_{F\sigma})$ is a dimensionless parameter controlling the value of the attenuation function $F(a_\sigma)$,
\begin{equation}
F(a_\sigma)=1-\frac{8}{3}a_\sigma\left[\sqrt{\pi}\text{erf}\left(\frac{1}{2a_\sigma}\right)-3a_\sigma+4a_\sigma^3+(2a_\sigma-4a_\sigma^3)\text{exp}\left(-\frac{1}{4a_\sigma^2}\right)\right].
\end{equation}

To develop a possible SR DFA exchange functional $E_x^{\text{SR-DFA}}$ based on the knowledge of a DFA exchange functional $E_x^{\text{DFA}}$, there are three schemes as follows. 
Consider the general expression of DFA exchange functional, which is   
\begin{equation}
E_x^{\text{DFA}}=\sum_\sigma \int e_{x\sigma}^\text{LDA}(\rho_\sigma)F_{x\sigma}^\text{DFA}d{\bf r},
\end{equation}
$F_{x\sigma}^\text{DFA}$ is the DFA enhancement factor for $\sigma$-spin. Depending on the type of DFA, $F_{x\sigma}^\text{DFA} = 1$ for a LDA, 
$F_{x\sigma}^\text{DFA} = F_{x\sigma}^\text{GGA}(\rho_\sigma,\nabla\rho_\sigma)$ for a GGA, $F_{x\sigma}^\text{DFA} = F_{x\sigma}^\text{MGGA}(\rho_\sigma,\nabla\rho_\sigma,\tau_\sigma)$ for a meta-GGA, 
where $\rho_{\sigma}({\bf r})$ is the spin density, $\nabla\rho_{\sigma}({\bf r})$ is the spin density gradient, and 
\begin{equation}
\tau_\sigma=\frac{1}{2}\sum_i^\text{occ.}\left|\nabla\psi_{i\sigma}\right|^2
\end{equation}
is the spin kinetic energy density.

The first scheme was proposed by Iikura, Tsuneda, Yanai, and Hirao (ITYH) \cite{Iikura01,Tawada04,Song07}, where $E_x^{\text{SR-DFA}}$ can be obtained by substituting a modified Fermi wave vector,
\begin{equation}
k_{\sigma}=\frac{k_{F\sigma}}{ \sqrt{ F_{x\sigma}^\text{DFA}} }
\end{equation}
into SR exchange energy density of Eq. (\ref{eSR_LDA}), which {\it a priori} produces $E_x^{\text{SR-DFA}}$ from any $E_x^{\text{DFA}}$, and reduces nicely to $E_x^{\text{SR-LDA}}$ from a $E_x^{\text{LDA}}$. 
Although the ITYH scheme possesses an admirable simplicity, some of its deficiencies (which potentially limit its accuracy) have been found \cite{Henderson08}. 

The second scheme was proposed by Vydrov, Heyd, Krukau, and Scuseria (VHKS) \cite{Vydrov06,Vydrov06_2}, where for a given spherically-averaged exchange hole $h_x^\text{DFA}({\bf r},r_{12})$, 
$E_x^{\text{SR-DFA}}$ is evaluated as
\begin{equation}
E_x^\text{SR-DFA}=2\pi\int\rho({\bf r})d{\bf r}\int_0^\infty\text{erfc}(\omega r_{12})h_x^\text{DFA}({\bf r},r_{12})r_{12}dr_{12}.
\end{equation}
The pivot of this scheme is the engineering of the DFA exchange hole. The GGA model exchange hole of Ernzerhof and Perdew \cite{EP} (EP) provides a framework for modeling any GGA exchange hole. 
It has made considerable appearances in real applications after parametrization to reproduce the Perdew, Burke, and Ernzerhof (PBE) GGA \cite{PBE, *PBE_E}. 
In 2008, Henderson, Janesko, and Scuseria \cite{Henderson08} (HJS) proposed another general model for the spherically averaged exchange hole corresponding to a GGA exchange functional, based on the work of EP. 
The HJS model improves upon the EP model by precisely reproducing the energy of the parent GGA, and by enabling fully analytic evaluation of range-separated hybrid density functionals. For meta-GGA, the TPSS exchange and correlation hole models have been ``reverse-engineered" \cite{Constantin06}. However, the resulting LC-TPSS functional (a LC hybrid MGGA) has no satisfactory long-range correction effect \cite{Vydrov06}.

The third scheme was proposed by Chai and Head-Gordon (CHG) \cite{Chai08,Chai08_2}, where $E_x^{\text{SR-DFA}}$ is evaluated as  
\begin{equation}
E_x^{\text{SR-DFA}}=\sum_\sigma\int e_{x\sigma}^\text{SR-LDA}(\rho_\sigma)F_{x\sigma}^\text{DFA}d{\bf r}.
\end{equation}
This simple scheme is expected to work well for a small $\omega$. For highly parametrized $E_x^{\text{DFA}}$, such as the B97 \cite{B97}, M05 \cite{Zhao05}, and M08 \cite{Zhao08} functionals, the CHG scheme 
is particularly attractive due to its simplicity. However, how large is not too large for the $\omega$ suitable for the CHG scheme? In the following sections, we will compare the performance of two new 
LC hybrid MGGA-D functionals, where one is developed by the CHG scheme, while the other is developed by a new scheme provided in this work, and our results help to answer the above question.

\section{LC hybrid MGGA-D functionals}

In this section, we introduce our new LC hybrid MGGA-D functionals. Note that LC-TPSS has been developed by utilizing the TPSS exchange hole (based on the VHKS scheme) \cite{Vydrov06}, but it is found that 
LC-TPSS does not benefit much by admixture of HF exchange. The M11 functional \cite{M11} has been developed based on the extension of a global hybrid MGGA functional, M08 \cite{Zhao08}, to LC-DFT, 
following the CHG scheme \cite{Chai08}. 

Parallel to the strategy of the $\omega$B97 series \cite{Chai08,Chai08_2}, we choose to modify the M05 functional. The M05 functional is a global hybrid MGGA functional with a powerful form \cite{Zhao05,Zhao06}, 
and our work is based on modifying this functional. Its exchange part consists of the PBE exchange functional and a reasonable kinetic-energy-density enhancement factor. The PBE exchange is a theoretically sound 
starting point because it satisfies the correct UEG limit and also has reasonable behavior at large values of the reduced spin density gradient $s_\sigma$.

To satisfy the UEG limit of SR exchange, we replace the PBE exchange energy density $e_{x\sigma}^\text{PBE}(\rho_\sigma,\nabla\rho_\sigma)$ with the SR-PBE exchange energy density 
$e_{x\sigma}^\text{SR-PBE}(\rho_\sigma,\nabla\rho_\sigma)$ generated by the HJS model exchange hole (based on the VHKS scheme), whose virtues are indicated in Sec. II. To achieve a flexible functional form, 
we retain the kinetic-energy-density enhancement factor (similar to the CHG scheme). 
We denote this resulting functional as SR-M05 (short-range M05) exchange, as it reduces to the M05 exchange at $\omega=0$.
\begin{equation}\label{ExSRM05}
E_x^\text{SR-M05}=\sum_\sigma\int e_{x\sigma}^\text{SR-PBE}(\rho_\sigma,\nabla\rho_\sigma)f(w_\sigma)d{\bf r},
\end{equation}
where $f(w_\sigma)$ is the kinetic-energy-density enhancement factor,
\begin{equation}\label{f}
f(w_\sigma)=\sum_{i=0}^ma_iw_\sigma^i.
\end{equation}
$w_\sigma$ is a function of $t_\sigma$, and $t_\sigma$ is a function of the kinetic energy density $\tau_\sigma$ of electrons with spin $\sigma$, as designed by Becke \cite{Becke00},
\begin{equation}w_\sigma=(t_\sigma-1)/(t_\sigma+1),\end{equation}
where
\begin{equation}t_\sigma=\tau_\sigma^\text{LDA}/\tau_\sigma,\end{equation}
\begin{equation}
\tau_\sigma^\text{LDA}\equiv\frac{3}{10}(6\pi^2)^{2/3}\rho_\sigma^{5/3}.
\end{equation}
In general, the enhancement factor should be $\omega$-dependent. But from the works of LC-TPSS \cite{Vydrov06} and M11 \cite{M11}, the optimal $\omega$ for a LC hybrid MGGA is expected to be small as well. 
For a sufficiently small $\omega$ value, our proposed functional form, inspired by the VHKS and CHG schemes, should be a good approximation. 

We use the same form for the correlation functional as the M05 correlation functional, which can be decomposed into same-spin $E_{c~\sigma\sigma}^\text{M05}$ and opposite-spin $E_{c~\alpha\beta}^\text{M05}$ components,
\begin{equation}
E_c^\text{M05}=\sum_\sigma E_{c~\sigma\sigma}^\text{M05}+E_{c~\alpha\beta}^\text{M05}.
\end{equation}
For the opposite-spin terms,
\begin{equation}
E_{c~\alpha\beta}^\text{M05}=\int e_{c~\alpha\beta}^\text{LDA}g_{\alpha\beta}(s_{av}^2)d{\bf r},
\end{equation}
\begin{equation}\label{gos}
g_{\alpha\beta}(s_{av}^2)=\sum_{i=0}^nc_{\alpha\beta,i}u_{\alpha\beta}^i,
\end{equation}
\begin{equation}
u_{\alpha\beta}=\frac{\gamma_{\alpha\beta}s_{av}^2}{1+\gamma_{\alpha\beta}s_{av}^2},
\end{equation}
\begin{equation}\gamma_{\alpha\beta}=0.0062,\end{equation}
\begin{equation}
s_{av}^2=\frac{1}{2}(s_\alpha^2+s_\beta^2),
\end{equation}
and for the same-spin terms,
\begin{equation}\label{ssC}
E_{c~\sigma\sigma}^\text{M05}=\int e_{c~\sigma\sigma}^\text{LDA}g_{\sigma\sigma}(s_\sigma^2)(1-\frac{\tau_\sigma^\text{W}}{\tau_\sigma})d{\bf r},
\end{equation}
\begin{equation}\label{gss}
g_{\sigma\sigma}(s_\sigma^2)=\sum_{i=0}^nc_{\sigma\sigma,i}u_{\sigma\sigma}^i,
\end{equation}
\begin{equation}
u_{\sigma\sigma}=\frac{\gamma_{\sigma\sigma}s_\sigma^2}{1+\gamma_{\sigma\sigma}s_\sigma^2},
\end{equation}
\begin{equation}\gamma_{\sigma\sigma}=0.06.\end{equation}
$1-\tau_\sigma^\text{W}/\tau_\sigma$ is a self-interaction correction factor proposed by Becke, \cite{Becke98} in which $\tau_\sigma^\text{W}$ is the von Weizsa$\ddot{\text{c}}$ker kinetic energy density \cite{von35} given by
\begin{equation}
\tau_\sigma^\text{W}=\frac{\left|\nabla\rho_\sigma\right|^2}{8\rho_\sigma}.
\end{equation}
In a one-electron case, $\tau_\sigma=\tau_\sigma^\text{W}$, so Eq. (\ref{ssC}) vanishes in any one-electron system. The correlation energy densities $e_{c~\alpha\beta}^\text{LDA}$ and $e_{c~\sigma\sigma}^\text{LDA}$ 
are derived from the Perdew-Wang parametrization of the LDA correlation energy \cite{Perdew92}, using the approach of Stoll {\it et al.} \cite{Stoll78,*Stoll80}, 
\begin{equation}
e_{c~\alpha\beta}^\text{LDA}(\rho_\alpha,\rho_\beta)=e_c^\text{LDA}(\rho_\alpha,\rho_\beta)-e_c^\text{LDA}(\rho_\alpha,0)-e_c^\text{LDA}(0,\rho_\beta),
\end{equation}
\begin{equation}e_{c~\sigma\sigma}^\text{LDA}=e_c^\text{LDA}(\rho_\sigma,0).\end{equation}

Based on the above functional expansions, we propose a new LC hybrid MGGA functional, $\omega$M05-D. It contains a fraction of the SR HF exchange,
\begin{equation}\label{Exc}
E_{xc}^{\omega\text{M05-D}}=E_x^\text{LR-HF}+c_xE_x^\text{SR-HF}+E_x^\text{SR-M05}+E_c^\text{M05}.
\end{equation}

We enforce the exact UEG limit for the $\omega$M05-D functional by imposing the following constraints:
\begin{equation}c_{\sigma\sigma,0}=1,\end{equation}
\begin{equation}c_{\alpha\beta,0}=1,\end{equation}
and
\begin{equation}a_0+c_x=1.\end{equation}

Following the general form of the DFT-D scheme \cite{Wu01,*Wu02,*Zimmerli04,*Grimme04,*Grimme06,*Antony06,*Jurecka06,*Goursot07,*Grimme07,*Cerny07,*Morgado07,*Kabelac07,*Cerny_PCCP07}, our total energy
\begin{equation}
E_\text{DFT-D}=E_\text{KS-DFT}+E_\text{disp}
\end{equation}
is computed as the sum of a KS-DFT part and an empirical atomic-pairwise dispersion correction. We choose to use the same form of unscaled dispersion correction as implemented in $\omega$B97X-D \cite{Chai08_2}, 
\begin{equation}
E_\text{disp}=-\sum_{i=1}^{N_\text{at}-1}\sum_{j=i+1}^{N_\text{at}}\frac{C_6^{ij}}{R_{ij}^6}f_\text{damp}(R_{ij}),
\end{equation}
where $N_\text{at}$ is the number of atoms in the system, $C_6^{ij}$ is the dispersion coefficient for atom pair $ij$, and $R_{ij}$ is an interatomic distance. The damping function,
\begin{equation}\label{fdamp}
f_\text{damp}(R_{ij})=\frac{1}{1+a(R_{ij}/R_r)^{-12}}
\end{equation}
enforces the conditions of zero dispersion correction at short interatomic separations and correct asymptotic pairwise vdW potentials. Here, $R_r$ is the sum of vdW radii of the atomic pair $ij$, and the only non-linear parameter, $a$, controls the strength of dispersion corrections.

To achieve an optimized functional for well-balanced performance across typical applications, we use the same diverse training set described in Ref. \cite{Chai08}, which contains 412 accurate experimental and accurate theoretical results, including the 18 atomic energies from the H atom to the Ar atom \cite{Chakravorty93}, the atomization energies of the G3/99 set (223 molecules) \cite{Curtiss97,Curtiss98,Curtiss00}, the ionization potentials (IPs) 
of the G2-1 set \cite{Pople89} (40 molecules, excluding \ce{SH2 (^2A1)} and \ce{N2 (^2\Pi)} cations due to the known convergence problems for semilocal density functionals \cite{Curtiss98}), the electron affinities (EAs) of the G2-1 set (25 molecules), 
the proton affinities (PAs) of the G2-1 set (8 molecules), the 76 barrier heights of the NHTBH38/04 and HTBH38/04 sets \cite{Zhao04,Zhao_JPCA05,*Zhao_E06}, and the 22 noncovalent interactions of the S22 
set \cite{Jurecka_PCCP06}. The S22 data is weighted ten times more than the others. All the parameters in $\omega$M05-D are determined self-consistently by a least-square fitting procedure described in Ref.\ \cite{Chai08}. 
For the non-linear parameter optimization, we focus on a range of possible $\omega$ values (0.0, 0.1, 0.2, 0.3, and 0.4 Bohr$^{-1}$), and optimize the corresponding $a$ values in the steps described in Ref.\ \cite{Chai08_2}.

M05 and M05-2X \cite{Zhao05,Zhao06} both used $m$=11 in Eq.\ (\ref{f}) and $n$=4 in Eqs.\ (\ref{gos}) and (\ref{gss}). However, during the optimization procedure of $\omega$M05-D, we found that the statistical errors are close 
for $m$=10 and $m$=11, while the one with $m$=11 has parameters significantly larger. A recent study by Wheeler and Houk has shown that large magnitude of the parameters in Eq. (\ref{f}) may result in large grid 
errors \cite{Wheeler10}. Moreover, the use of large parameters increases the possibility of convergence difficulty as well as the over-fitting effects. Thus, we choose $m$=10 instead of 11 in Eq. (\ref{f}). The optimized parameters of 
the $\omega$M05-D functional are given in Table \ref{para}, in which the $\omega$ value is same as that of $\omega$B97X-D, while the fraction of SR HF exchange, $c_x$, is larger than that of $\omega$B97X-D ($\approx$ 0.22). 
This helps to reduce the self-interaction error (SIE) of the functional, as can be seen in Sec. V.

We also tried a simple model (based on the CHG scheme), where the SR-PBE exchange energy density $e_{x\sigma}^\text{SR-PBE}(\rho_\sigma,\nabla\rho_\sigma)$ used in Eq. (\ref{ExSRM05}) is substituted with 
$e_{x\sigma}^\text{SR-LDA}(\rho_\sigma)F_x^\text{PBE}(s_\sigma)$, that is, the SR LDA exchange energy density in Eq. (\ref{eSR_LDA}) multiplied by the PBE enhancement factor. We tried this because the 
mathematical form of the latter is significantly simpler than that of the former, and is the model on which M11 based. The parametrization is the same for this simple model, which we denoted by $\omega$M05s-D. 
Compared to $\omega$M05-D, the optimal $\omega$ value is also 0.2 bohr$^{-1}$, but the corresponding optimal $a$ value is found to be 100 and the linear parameters are also larger.

\begin{table}
\caption{\label{para}Optimized parameters for $\omega $M05-D. Here, the non-linear parameter $a$ is defined in Eq. (\ref{fdamp}), and others are defined in Eq. (\ref{Exc})}
\begin{ruledtabular}
\begin{tabular}{cccc}
\textit{a} & \multicolumn{3}{c}{30.0} \\ 
\textit{$\omega $} & \multicolumn{3}{c}{0.2 Bohr${}^{-1}$} \\
\textit{c${}_{x}$} & \multicolumn{3}{c}{0.369592} \\ \hline
\textit{i} & $a_i$ & $c_{\alpha \beta ,i}$ & $c_{\sigma \sigma ,i}$ \\ \hline 
0 & 0.630408 & 1.00000 & 1.00000 \\
1 & -0.219121 & -0.95491 & -5.26863 \\ 
2 & -0.14411 & 12.138 & 17.9935 \\ 
3 & 1.27732 & -35.1041 & -17.6408 \\
4 & -1.59959 & 19.5804 & 0.625687 \\
5 & -5.94702 &  &  \\
6 & 13.5822 &  &  \\
7 & 10.5048 &  &  \\
8 & -28.7168 &  &  \\ 
9 & -6.89761 &  &  \\
10 & 19.0574 &  &  \\
\end{tabular}
\end{ruledtabular}
\end{table}

\section{RESULTS FOR THE TRAINING SET}

All calculations are performed with a development version of \textsf{Q-CHEM 3.2} \cite{QChem3}. Spin-restricted theory is used for singlet states and spin-unrestricted theory for others, unless noted otherwise. 
For the binding energies of the weakly bound systems, the counterpoise correction \cite{Boys70} is employed to reduce basis set superposition error (BSSE).

Results for the training set are computed using the 6-311++G(3df,3pd) basis set with the fine grid, EML(75,302), consisting of 75 Euler-Maclaurin radial grid points \cite{Murray93} and 302 Lebedev angular grid 
points \cite{Lebedev75,*Lebedev76,*Lebedev77}. The error for each entry is defined as error = theoretical value $-$ reference value. The notation used for characterizing statistical errors is as follows: mean signed 
errors (MSEs), mean absolute errors (MAEs), root-mean-square (rms) errors, maximum negative errors (Max($-$)), and maximum positive errors (Max(+)).

First, we show the results of the first iteration of fitting procedure, comparing the new LC scheme with the CHG scheme (the simple model) for $\omega$=0.1, 0.2, 0.3 and 0.4 bohr$^{-1}$. We optimize $\omega$M05 
and $\omega$M05s using the corresponding $\omega$PBE and $\omega$PBEs orbitals, and denote these optimized functionals as $\omega$M05* and $\omega$M05s*. The statistical errors are believed to be quite close to those obtained self-consistently. As can be seen in Table \ref{schemes}, the difference between the performance of $\omega$M05* and $\omega$M05s* is noticeable for $\omega$=0.2 bohr$^{-1}$, and becomes larger for 
a larger $\omega$ value. Therefore, a LC hybrid MGGA functional with a larger $\omega$ value (such as M11 with $\omega$=0.25 bohr$^{-1}$) may perform better with our new scheme than with the CHG scheme.

In subsequent iterations, we include the dispersion corrections, increase the training weight of S22 set, and found the functionals optimized with $\omega$=0.2 bohr$^{-1}$.
To view the effect of the long-range correction and the dispersion corrections, we also consider the functional form M05 and M05-D. The latter is the limiting case where $\omega$=0 for $\omega$M05-D, of which the corresponding optimal $a$ value is found to be 2. We reoptimize M05 and M05-D functionals on the same training set using the M05-2X orbitals, truncate their functional expansions at the same orders $m$=10 and $n$=4, and denote these two reoptimized functionals as M05* and M05-D*. Just like the $\omega$B97X functional without dispersion correction, all data in the training set are equally weighted in the least-squares fitting for M05*.

The overall performance of our new $\omega$M05-D is compared with the trial simple model $\omega$M05s-D, M05-D*, M05* and M05-2X \cite{Zhao06}, as well as existing $\omega$B97X-D (a LC hybrid GGA-D) \cite{Chai08_2}. 
Note that M05 \cite{Zhao05} and M05-2X share the same functional form, but the former is distracted to deal with transition-metal compounds, so the latter should be our concern. In the $\omega$B97 series, $\omega$B97X-D has the closest relationship to $\omega$M05-D, while $\omega$B97 and $\omega$B97X, developed without dispersion corrections, are expected to perform poorly for noncovalent interactions.

In Table \ref{train}, the first comparison ($\omega$M05-D vs. $\omega$M05s-D) partially determines the choice of our proposed functional. Although $\omega$M05-D performs worse than $\omega$M05s-D for HTBH, the overall performance of $\omega$M05-D in the training set is the best. 

A second comparison between $\omega$M05-D and M05-D* indicates that the exact long-range exchange indeed leads to an overall improvement to MGGA, although not as large as that to GGA \cite{Vydrov06,Chai08,Chai08_2}. 
The third comparison is between M05-D* and M05*. The cooperation of the training weight and the empirical dispersion corrections leads to a significant improvement in the results for noncovalent interactions (the S22 data) and a modest overall change. Recently, there have been the updated reference values for the S22 set \cite{Marshall11}. We have also examined the performance of $\omega$M05-D against the updated S22 reference values. 
As shown in the supplementary material \cite{SI}, the overall performance of the functional against the updated reference values is similar to that against the original ones. 

\begin{table}
\caption{\label{schemes}Comparisons between the $\omega$M05* and $\omega$M05s* functionals (defined in the text) for different $\omega$ values. Statistical errors are in kcal/mol.}
\begin{ruledtabular}
\begin{tabular}{ccrrrrrrrr} 
	&	$\omega$ (Bohr$^{-1}$)	&	0.1	&	0.1	&	0.2	&	0.2	&	0.3	&	0.3	&	0.4	&	0.4	\\
System	&	Error	&	$\omega$M05*	&	$\omega$M05s*	&	$\omega$M05*	&	$\omega$M05s*	&	$\omega$M05*	&	$\omega$M05s*	&	$\omega$M05*	&	$\omega$M05s*	\\ \hline 
Atoms	&	MSE	&	-0.15 	&	0.24 	&	0.05 	&	0.63 	&	0.23 	&	0.97 	&	0.46 	&	1.33 	\\
(18)	&	MAE	&	2.02 	&	2.09 	&	1.81 	&	2.35 	&	2.00 	&	3.36 	&	3.22 	&	5.05 	\\
	&		&		&		&		&		&		&		&		&		\\
G3/99	&	MSE	&	0.06 	&	0.09 	&	0.05 	&	0.03 	&	-0.04 	&	-0.12 	&	-0.18 	&	-0.27 	\\
(223)	&	MAE	&	1.77 	&	1.79 	&	1.66 	&	1.76 	&	1.78 	&	2.02 	&	2.10 	&	2.35 	\\
	&		&		&		&		&		&		&		&		&		\\
IP	&	MSE	&	-0.58 	&	-1.48 	&	-0.84 	&	-1.36 	&	-0.38 	&	-0.32 	&	0.30 	&	0.73 	\\
(40)	&	MAE	&	2.75 	&	3.06 	&	2.81 	&	3.08 	&	2.68 	&	2.81 	&	2.64 	&	2.79 	\\
	&		&		&		&		&		&		&		&		&		\\
EA	&	MSE	&	-1.50 	&	-1.70 	&	-1.29 	&	-1.15 	&	-0.94 	&	-0.70 	&	-0.64 	&	-0.39 	\\
(25)	&	MAE	&	2.50 	&	2.56 	&	2.33 	&	2.22 	&	2.07 	&	1.97 	&	1.98 	&	1.91 	\\
	&		&		&		&		&		&		&		&		&		\\
PA	&	MSE	&	-1.65 	&	-2.68 	&	-1.49 	&	-2.71 	&	-1.07 	&	-2.11 	&	-0.78 	&	-1.54 	\\
(8)	&	MAE	&	1.87 	&	2.68 	&	1.83 	&	2.82 	&	1.79 	&	2.53 	&	1.86 	&	2.43 	\\
	&		&		&		&		&		&		&		&		&		\\
NHTBH	&	MSE	&	-1.26 	&	-1.09 	&	-0.68 	&	-0.39 	&	0.08 	&	0.40 	&	0.85 	&	1.17 	\\
(38)	&	MAE	&	1.98 	&	1.82 	&	1.51 	&	1.40 	&	1.46 	&	1.67 	&	1.71 	&	1.95 	\\
	&		&		&		&		&		&		&		&		&		\\
HTBH	&	MSE	&	-1.96 	&	-1.95 	&	-1.95 	&	-1.68 	&	-1.61 	&	-1.24 	&	-1.21 	&	-0.84 	\\
(38)	&	MAE	&	2.19 	&	2.08 	&	2.12 	&	1.84 	&	1.95 	&	1.56 	&	1.86 	&	1.47 	\\
	&		&		&		&		&		&		&		&		&		\\
S22	&	MSE	&	2.65 	&	1.91 	&	1.80 	&	1.06 	&	1.01 	&	0.56 	&	0.45 	&	0.26 	\\
(22)	&	MAE	&	2.65 	&	1.91 	&	1.80 	&	1.06 	&	1.02 	&	0.67 	&	0.71 	&	0.63 	\\
	&		&		&		&		&		&		&		&		&		\\
All	&	MSE	&	-0.31 	&	-0.42 	&	-0.30 	&	-0.35 	&	-0.21 	&	-0.18 	&	-0.11 	&	-0.02 	\\
(412)	&	MAE	&	2.03 	&	2.03 	&	1.86 	&	1.90 	&	1.84 	&	2.01 	&	2.06 	&	2.28 	\\
\end{tabular}
\end{ruledtabular}
\end{table}

\begin{table}
\caption{\label{train}Statistical errors (in kcal/mol) of the training set. The M05-D* and M05* functionals are defined in the text. M05-2X was not particularly parametrized using this training set.}
\begin{ruledtabular}
\begin{tabular}{ccrrrrrr} 
System & Error & $\omega$M05-D & $\omega$M05s-D & M05-D* & M05* &  M05-2X & $\omega$B97X-D  \\ \hline 
Atoms  & MSE & 0.37 & 0.83 & 0.18  & -0.48 &  -3.01 & -0.05  \\
(18) & MAE & 2.02 & 2.28 & 2.61  & 1.98 &  5.10 & 2.57  \\
 &  &  &  &  &  &  &    \\
G3/99  & MSE & -0.03 & -0.03   & -0.10& -0.05 &  2.01 & -0.24  \\
(223) & MAE & 1.62 & 1.73 & 1.78  & 1.78 &  3.65 & 1.93  \\ 
 &  &  &  &  &  &  &    \\
IP  & MSE & -0.80 & -1.33 & 0.06  & 0.27 &  1.10 & 0.19  \\
(40) & MAE & 2.86 & 3.04 & 2.84  & 2.51 &  3.35 & 2.74  \\
 &  &  &  &  &  &  &    \\
EA  & MSE & -1.02 & -0.98 & -0.54  & -0.84 &  -0.23 & 0.07  \\ 
(25) & MAE & 2.12 & 2.13 & 2.13  & 2.35 &  2.48 & 1.91  \\
 &  &  &  &  &  &  &    \\
PA  & MSE & -1.48 & -2.66 & -0.94  & -1.76 &  -1.26 & 1.42  \\
(8) & MAE & 2.10 & 3.07 & 1.31  & 2.17 &  1.51 & 1.50  \\
 &  &  &  &  &  &  &    \\
NHTBH  & MSE & -0.94 & -0.59 & -1.38  & -1.32 &  0.13 & -0.45  \\
(38) & MAE & 1.57 & 1.53 & 2.04 & 2.08 &  1.75 & 1.51  \\
 &  &  &  &  &  &  &    \\
HTBH  & MSE & -2.82 & -2.33 & -2.95 & -1.77 &  -0.65 & -2.57  \\
(38) & MAE & 2.83 & 2.37 & 3.08 & 2.14 &  1.51 & 2.70  \\
 &  &  &  &  &  &  &    \\
S22  & MSE & -0.01 & -0.01 & 0.04 & 3.46 &  0.73 & -0.08  \\
(22) & MAE & 0.27 & 0.21 & 0.23 & 3.46 &  0.87 & 0.21  \\
 &  &  &  &  &  &  &    \\
All  & MSE & -0.51 & -0.49 & -0.49 & -0.21 &  1.02 & -0.36  \\
(412) & MAE & 1.83 & 1.89 & 1.99 & 2.05 &  3.05 & 1.96  \\
\end{tabular}
\end{ruledtabular}
\end{table}

\section{RESULTS FOR THE TEST SETS}

To test the performance of $\omega$M05-D outside its training set, we also evaluate its performance on various test sets involving 48 atomization energies in the G3/05 test set \cite{Curtiss05}, 30 chemical reaction energies 
taken from the NHTBH38/04 and HTBH38/04 databases \cite{Zhao04,Zhao_JPCA05,*Zhao_E06}, 29 noncovalent interactions \cite{Zhao_JPCA05,Jurecka_PCCP06}, 166 optimized geometry properties of covalent 
systems \cite{DiStasio07}, 12 intermolecular bond lengths \cite{Jurecka_PCCP06}, 4 dissociation curves of symmetric radical cations as well as three new databases, consisting of 131 vertical IPs, 115 vertical EAs and 115 fundamental gaps. For excitation energies, we perform TDDFT calculations for 19 valence excitation energies, 23 Rydberg excitation energies and one long-range charge transfer excitation curve of two well-separated molecules. Each EA can be evaluated by two different ways, and each fundamental gap can be evaluated by three different ways, so there are a total of 1038 pieces of data in the test sets, which are larger and more diverse than the training set. Unspecified detailed information of the test sets as well as the basis sets, and numerical grids used is given in Ref.\ \cite{Chai08}.

\subsection{Atomization Energies, Reaction Energies, and Noncovalent Interactions}

Table \ref{test} summarized the general energetic results in the same way as in Ref.\ \cite{Chai08_2}, for convenience of further comparisons. Since the 30 chemical reaction energies are taken from the NHTBH38/04 and 
HTBH38/04 databases calculated in Table \ref{train}, the EML(75,302) grid is used. In Table \ref{test}, the comparison between $\omega$M05-D and $\omega$M05s-D shows noticeable difference in atomization energies, 
and makes great influence on the choice of our proposed functional.

\begin{table}
\caption{\label{test} Statistical errors (in kcal/mol) of the test sets.}
\begin{ruledtabular}
\begin{tabular}{ccrrrrr}
System & Error & \textit{$\omega$}M05-D & $\omega$M05s-D &  M05-2X & $\omega$B97X-D \\ \hline
G3/05& MSE & -0.85 & -1.67 & 0.00 & 0.25 \\
(48) & MAE & 3.21 & 3.79 &  5.24 & 3.02 \\
 &  &  &  &    \\
RE & MSE & -0.58 & -0.65 &  -0.86 & -0.24 \\ 
(30) & MAE & 1.49 & 1.32 &  1.65 & 1.63 \\
 &  &  &  &    \\
Non-covalent & MSE & -0.11 & -0.05 &  0.50 & -0.15 \\ 
(29) & MAE & 0.31 & 0.30 & 0.61 & 0.43 \\
 &  &  &  &    \\
All & MSE & -0.58  & -0.95  &  -0.11  & 0.01  \\ 
(107) & MAE & 1.94  & 2.15 &  2.98  & 1.93 \\
\end{tabular}
\end{ruledtabular}
\end{table}

\subsection{Equilibrium Geometries}

Satisfactory predictions of molecular geometries of covalent and non-covalent systems by density functionals are necessary for practical use. For covalent systems, we perform geometry optimizations for each functional on 
the equilibrium experimental test set (EXTS) \cite{DiStasio07}, while for non-covalent systems, we compute the intermolecular bond lengths of 12 weakly bound complexes taken from the S22 set \cite{Jurecka_PCCP06}, 
using 6-311++G(3df,3pd) basis set with the EML(75,302) grid. As shown in Table \ref{geo}, performance of all the hybrid functionals in predicting optimized geometries of EXTS is similar, while the performance of 
simple model ($\omega$M05s-D) is somewhat worse for the intermolecular bond lengths. We decide our proposed model to be $\omega$M05-D in this subsection. For brevity, the performance of 
$\omega$M05s-D will not be shown for subsequent calculations.

\begin{table}
\caption{\label{geo}Statistical errors (in $\text{\AA}$) of EXTS (Ref. \cite{DiStasio07}) and bond lengths of 12 weakly bound complexes from the S22 set (Ref. \cite{Jurecka_PCCP06}). The results of
\textit{$\omega$}B97X-D are taken from Ref. \cite{Chai08_2}.}
\begin{ruledtabular}
\begin{tabular}{llrrrr}
System & Error & \textit{$\omega$}M05-D & $\omega$M05s-D & M05-2X & $\omega$B97X-D \\ \hline
 & MSE & 0.003 & 0.001 & -0.004 & -0.002 \\
 & MAE & 0.010 & 0.009 & 0.009 & 0.009 \\
EXTS (166) & rms & 0.019 & 0.014 & 0.014 & 0.013 \\
 & Max($-$) & -0.081 & -0.083 & -0.082 & -0.078 \\
 & Max(+) & 0.177 & 0.067 & 0.054 & 0.055 \\ \hline
 & MSE & -0.041  & -0.069  & -0.021 & -0.044  \\
 & MAE & 0.061  & 0.078  & 0.062 & 0.064  \\
Weak (12) & rms & 0.083  & 0.102  & 0.080 & 0.085  \\
 & Max($-$) & -0.189  & -0.195  & -0.165 & -0.198  \\
 & Max(+) & 0.043  & 0.029  & 0.140 & 0.056  \\
\end{tabular}
\end{ruledtabular}
\end{table}

\subsection{Dissociation of Symmetric Radical Cations}

Common semilocal functionals are generally accurate for systems near equilibrium. However, due to considerable self-interaction errors in semilocal functionals, spurious fractional charge dissociation 
occurs \cite{Bally97,*Braida98,*MoriS06,*Ruzsinszky07,Dutoi06,Vydrov06_2}. This situation becomes amplified for symmetric charged radicals \ce{X2+}, such as \ce{H2+}, \ce{He2+}, \ce{Ne2+} and \ce{Ar2+}. Gr$\ddot{\text{a}}$fenstein and coworkers have obtained qualitatively correct result for these systems \cite{Grafenstein_JCP04,Grafenstein04} using self-interaction-corrected DFT proposed by Perdew and Zunger \cite{PZ}, and confirmed that 
the errors of standard DFT methods should be dominated by the SIEs.

\begin{figure}
\includegraphics[width=15cm]{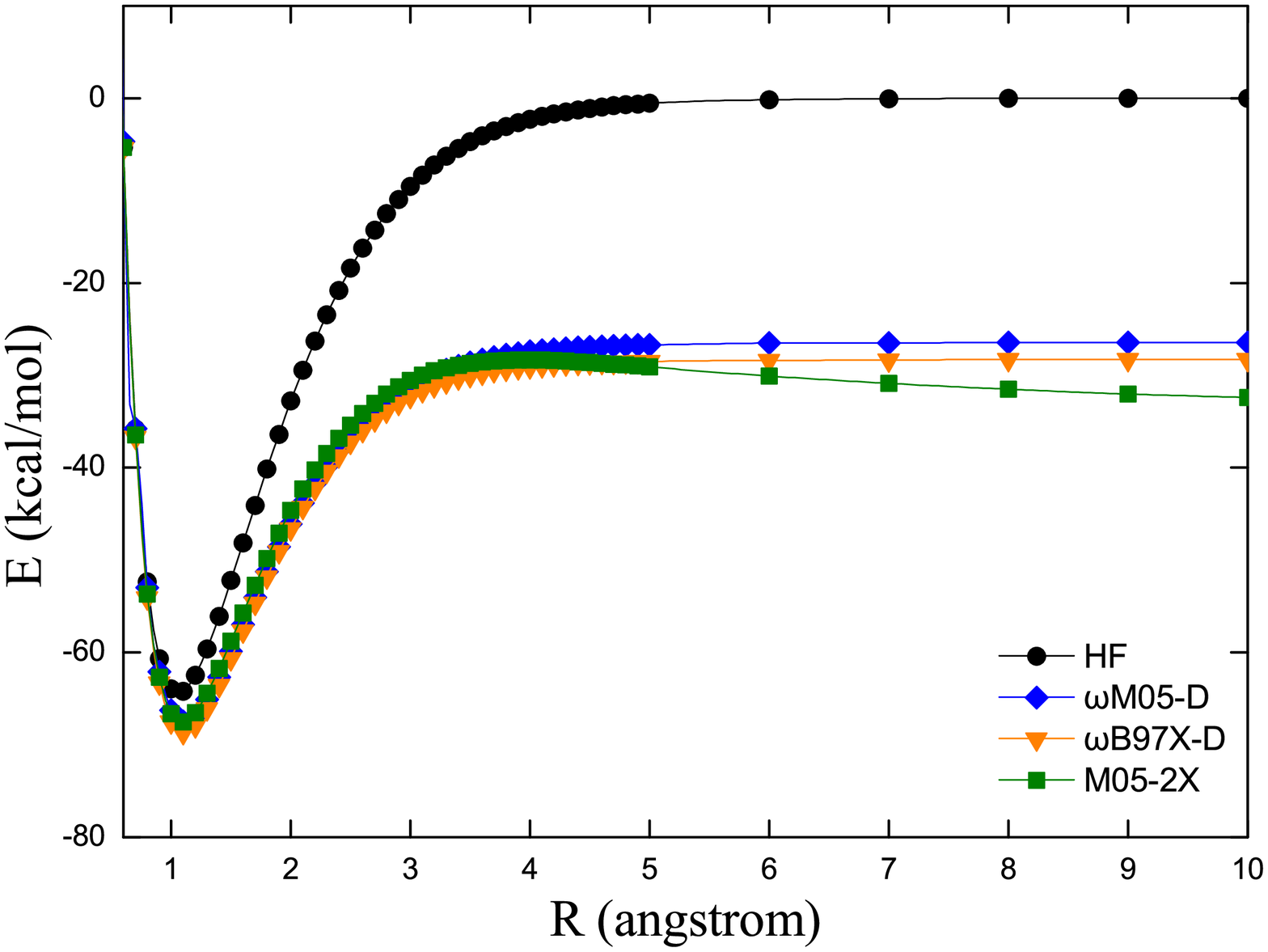}
\caption{Dissociation curve of \ce{H2+}. Zero level is set to E(H)+E(\ce{H+}) for each method.}
\label{H2}
\end{figure}

\begin{figure}
\includegraphics[width=15cm]{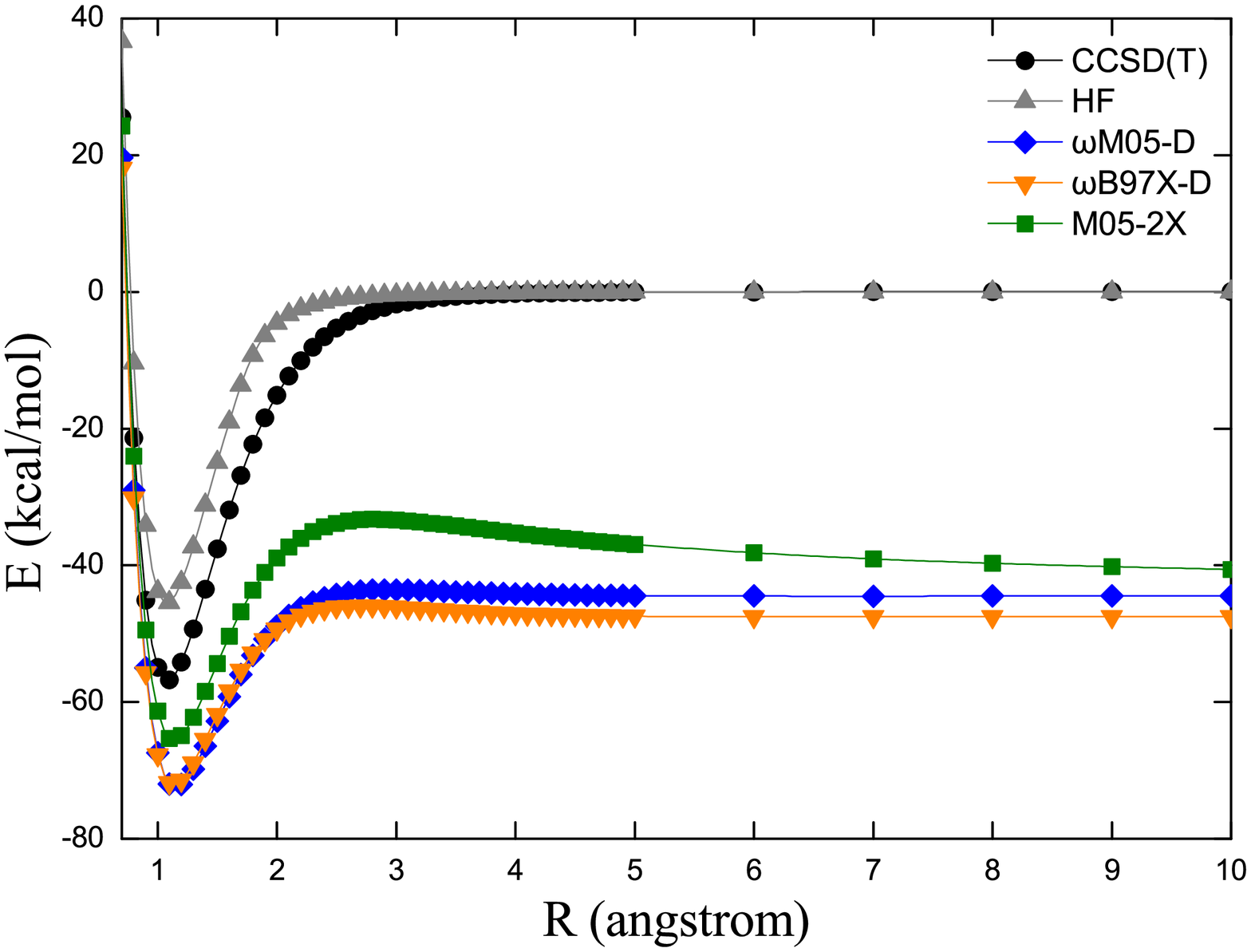}
\caption{Dissociation curve of \ce{He2+}. Zero level is set to E(He)+E(\ce{He+}) for each method.}
\label{He2}
\end{figure}

\begin{figure}
\includegraphics[width=15cm]{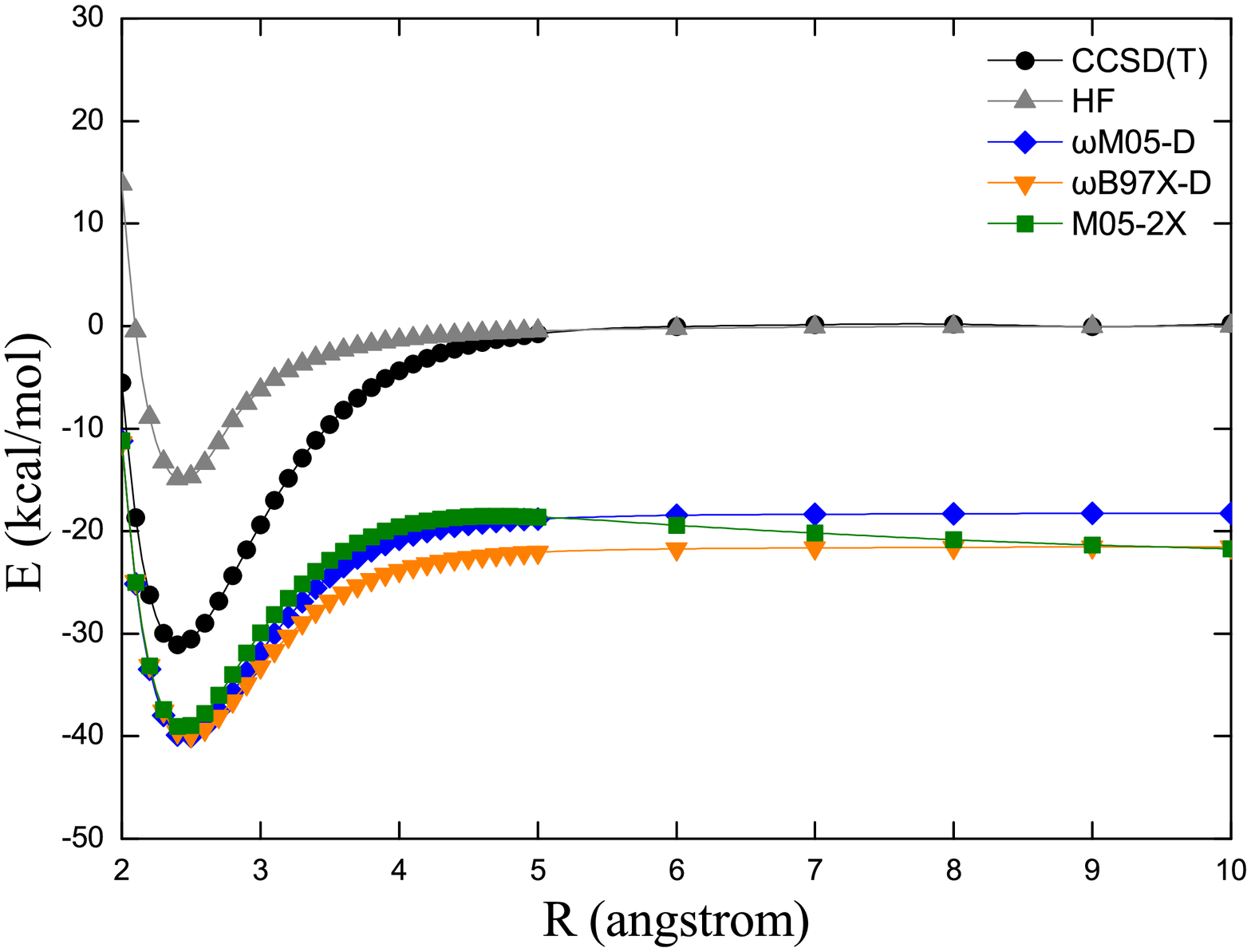}
\caption{Dissociation curve of \ce{Ar2+}. Zero level is set to E(Ar)+E(\ce{Ar+}) for each method.}
\label{Ar2}
\end{figure}

We perform unrestricted calculations with the aug-cc-pVQZ basis set and a high-quality EML(250,590) grid. The DFT results are compared with results from HF theory, and the very accurate CCSD(T) theory \cite{Purvis82,Raghavachari89}. The HF method is exact in Fig.\ \ref{H2}, and gives qualitatively correct results from Fig.\ \ref{He2} to Fig. \ref{Ar2}. Although $\omega$M05-D has the same amount of LR HF exchange 
as $\omega$B97X-D, the larger fraction of SR HF exchange included in $\omega$M05-D helps to reduce its remaining SIE. Therefore, the error of $\omega$M05-D is smaller than that of $\omega$B97X-D, especially for 
larger cations (e.g.\ \ce{Ne2+} and \ce{Ar2+}). The global hybrid functional M05-2X exhibits the undesirable \ce{X2+} dissociation curves, displaying a spurious energy barrier at intermediate bond length $R$.

\begin{figure}
\includegraphics[width=15cm]{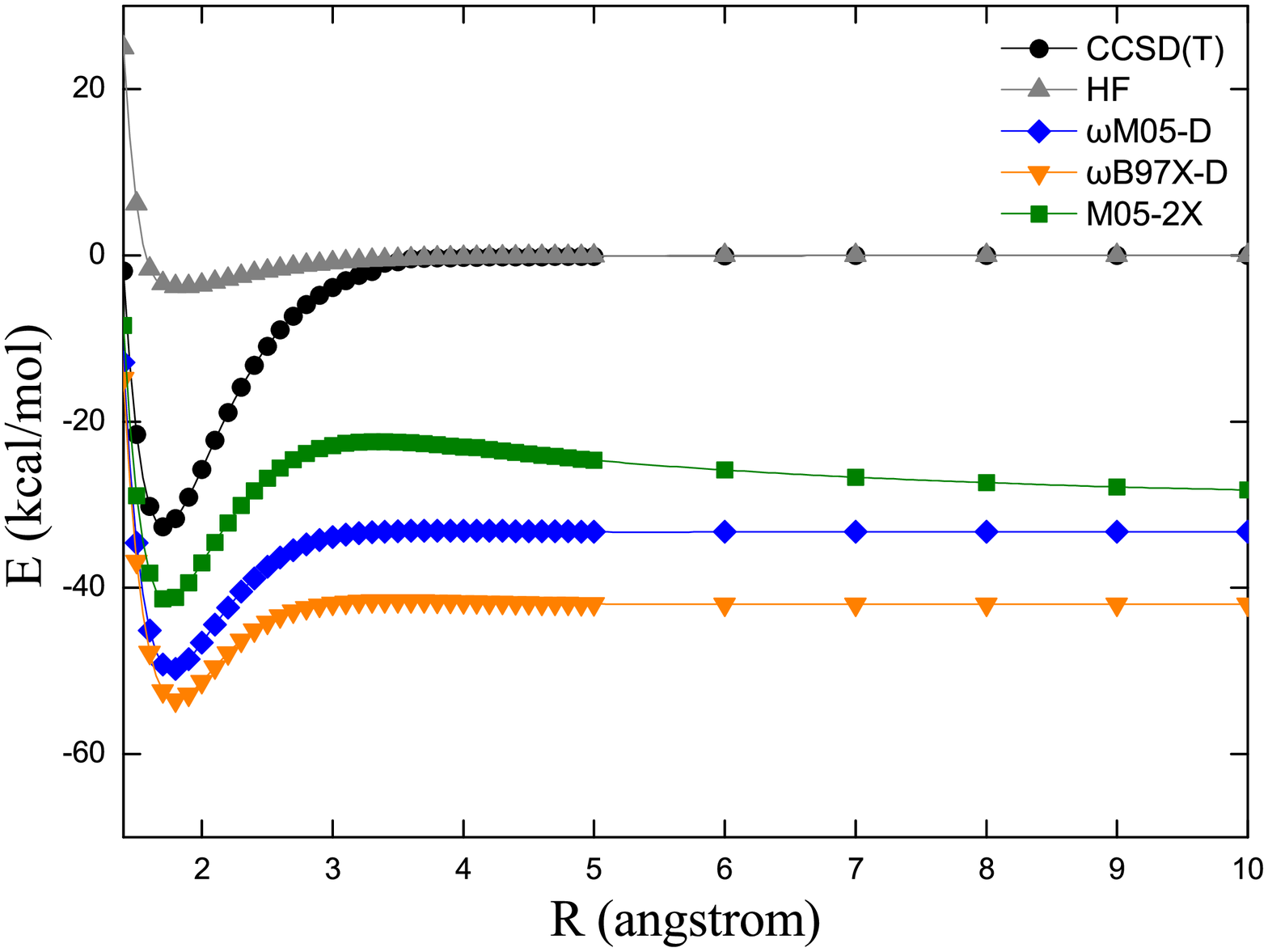}
\caption{Dissociation curve of \ce{Ne2+}. Zero level is set to E(Ne)+E(\ce{Ne+}) for each method.}
\end{figure}

\subsection{Frontier Orbital Energies}

Let IP($N$) be the ionization potential and EA($N$) be the electron affinity of the $N$-electron system, which are defined as
\begin{equation}\label{defIP}
\text{IP}(N) = {E}_{N-1} - {E}_{N},
\end{equation}
\begin{equation}\label{defEA}
\text{EA}(N) =  {E}_{N} - {E}_{N+1},
\end{equation}
respectively, with ${E}_{N}$ being the total energy of $N$-electron system. For the exact DFT, the vertical ionization potential of a neutral molecule is identical to the minus HOMO (highest occupied molecular orbital) energy of the neutral molecule \cite{Levy84, Parr89}, 
\begin{equation}\label{eqIP}
\text{IP}(N) = -{\epsilon}_{N}(N),
\end{equation}
and the vertical electron affinity of a neutral molecule is identical to the minus HOMO energy of the anion (since EA($N$) = IP($N+1$) by definition),
\begin{equation}\label{eqEA}
\text{EA}(N) = -{\epsilon}_{N+1}(N+1),
\end{equation}
where ${\epsilon}_{M}(N)$ is the $M$-th orbital energy of $N$-electron system. The vertical electron affinity of a neutral molecule may also be approximated by the minus LUMO (lowest unoccupied molecular orbital) energy 
of the neutral molecule, but it is proved that there exists a difference between the vertical EA and the minus LUMO energy,
\begin{equation}\label{delxc}
{\Delta}_{xc}={\epsilon}_{N+1}(N+1)-{\epsilon}_{N+1}(N),
\end{equation}
where the difference ${\Delta}_{xc}$ arises from the discontinuity of exchange-correlation potentials \cite{Sham83,Sham85,Perdew83}. Recent study shows that ${\Delta}_{xc}$ is close to zero for LC hybrid functionals \cite{Tsuneda10}, so the minus LUMO energy calculated by a LC hybrid functional should be close to the vertical EA.

To evaluate the performance of the functionals on the HOMO energy of the neutral molecule, we collect a new database, IP131, which consists of experimental vertical IPs of 18 atoms and 113 molecules in the experimental
geometries. The geometries and most of the reference values are collected from the NIST database \cite{NIST}. Other publications \cite{Gill92,*Brundle72,*Niessen82,*Colbourne78,*Frost72,*Bieri82,*Cvitas77,*Bieri81,*Asbrink80,*Gelius71,*Asbrink81,*Bieri80,*Niessen80} are adopted for the experimental vertical IPs of some molecules. The DFT calculations are performed with 6-311++G(3df,3pd) basis and EML(75,302) grid. As can be seen in Table \ref{IP}, $\omega$M05-D gives the best results. The global hybrid M05-2X gives the worst results here, due to its incorrect long-range XC-potential behavior.

To evaluate the performance of the functionals on the vertical electron affinity, we construct another database called EA115, which consists of 18 atoms and 97 molecules. For the molecular geometries, it is a subset of IP131. Because experimental vertical EAs are not as widely available as experimental vertical IPs, the reference values of vertical EAs are obtained via the accurate CCSD(T) calculations (using Eq.\ (\ref{defEA})). The CCSD(T) correlation energies in the basis-set limit are extrapolated from calculations using the aug-cc-pVTZ and aug-cc-pVQZ basis sets \cite{Halkier98}:
\begin{equation}
E_{XY}^\infty = \frac{E_X^\text{corr}X^3-E_Y^\text{corr}Y^3}{X^3-Y^3},
\end{equation}
where $X$=3 and $Y$=4 for the aug-cc-pVTZ and aug-cc-pVQZ basis, respectively. The electron affinities are evaluated in two different ways, as shown in Table \ref{EA} for the minus HOMO energy of the anion, and Table \ref{EA_LUMO} for the minus LUMO energy of the neutral molecule. Clearly, the LC hybrid functionals outperform the global hybrid M05-2X. The reference values and molecular geometries of IP131 and EA115 are given in the supplementary material \cite{SI} along with detailed DFT results.

\begin{table}
\caption{\label{IP}Statistical errors (in eV) for the IP131 database. Error is defined as - $\epsilon_N(N)$ - IP$_\text{vertical}$. Experimental geometries and reference values are used for all molecules.}
\begin{ruledtabular}
\begin{tabular}{ccccc}
System & Error & \textit{$\omega $}M05-D & M05-2X & \textit{$\omega $}B97X-D \\ \hline 
atoms & MSE & -1.48 & -2.06 & -1.64 \\ 
(18) & MAE & 1.48 & 2.06 & 1.64 \\ 
 & rms & 1.74 & 2.16 & 1.98 \\ \hline
molecules & MSE & -0.68 & -1.23 & -0.92 \\ 
(113) & MAE & 0.68 & 1.23 & 0.92 \\ 
 & rms & 0.76 & 1.27 & 1.00 \\ \hline
total & MSE & -0.79 & -1.34 & -1.02 \\ 
(131) & MAE & 0.79 & 1.34 & 1.02 \\ 
 & rms & 0.96 & 1.43 & 1.18 \\ 
\end{tabular}
\end{ruledtabular}
\end{table}

\begin{table}
\caption{\label{EA}Statistical errors (in eV) for the EA115 database. Error is defined as - $\epsilon_{N+1}(N+1)$ - EA$_\text{vertical}$. Experimental geometries and CCSD(T) reference values are used for all molecules.}
\begin{ruledtabular}
\begin{tabular}{ccccc}
System	&	Error	&	$\omega$M05-D	&	M05-2X	&	$\omega$B97X-D	\\	\hline
atoms	&	MSE	&	-0.46 	&	-1.21 	&	-0.53 	\\	
(18)	&	MAE	&	0.49 	&	1.21 	&	0.57 	\\	
	&	rms	&	0.73 	&	1.35 	&	0.84 	\\	\hline
moelcules	&	MSE	&	-0.54 	&	-1.18 	&	-0.54 	\\	
(97)	&	MAE	&	0.55 	&	1.18 	&	0.56 	\\	
	&	rms	&	0.80 	&	1.32 	&	0.82 	\\	\hline
total	&	MSE	&	-0.53 	&	-1.18 	&	-0.54 	\\	
(115)	&	MAE	&	0.55 	&	1.18 	&	0.56 	\\	
	&	rms	&	0.79 	&	1.32 	&	0.82 	\\	
\end{tabular}
\end{ruledtabular}
\end{table}

\begin{table}
\caption{\label{EA_LUMO}Statistical errors (in eV) of the minus LUMO energy of the neutral
molecule for the EA115 database. Experimental geometries and CCSD(T) reference values are used for all molecules.}
\begin{ruledtabular}
\begin{tabular}{ccccc}
System	&	Error	&	$\omega$M05-D	&	M05-2X	&	$\omega$B97X-D	\\	\hline
atoms	&	MSE	&	-0.27 	&	0.57 	&	-0.02 	\\	
(18)	&	MAE	&	0.73 	&	1.02 	&	0.74 	\\	
	&	rms	&	0.92 	&	1.12 	&	0.89 	\\	\hline
moelcules	&	MSE	&	-0.24 	&	0.60 	&	0.05 	\\	
(97)	&	MAE	&	0.60 	&	0.75 	&	0.52 	\\	
	&	rms	&	0.69 	&	0.94 	&	0.60 	\\	\hline
total	&	MSE	&	-0.24 	&	0.60 	&	0.04 	\\	
(115)	&	MAE	&	0.62 	&	0.79 	&	0.55 	\\	
	&	rms	&	0.73 	&	0.97 	&	0.65 	\\	
\end{tabular}
\end{ruledtabular}
\end{table}

\subsection{Fundamental Gaps}

The fundamental gap $E_{g}$ of a molecule with $N$ electrons is defined as
\begin{equation}
E_{g} = \text{IP}(N) - \text{EA}(N).
\label{eq:FBG}
\end{equation}
Following Eqs.\ (\ref{defIP}) and (\ref{defEA}) for the definitions of IP and EA, three self-consistent field (SCF) calculations (for the neutral molecule, cation and anion) are required to obtain the fundamental gap of a molecule. Using Eq. (\ref{eqIP}) and (\ref{eqEA}), the fundamental gap of a molecule can also be obtained by two SCF calculations (for the neutral molecule and anion).

Following Janak's theorem \cite{Janak78}, the fundamental gap can be approximated by the HOMO-LUMO gap \cite{Perdew83}
\begin{equation}
{\Delta}_{KS} = {\epsilon}_{N+1}(N) - {\epsilon}_{N}(N),
\label{eq:HOMO-LUMO}
\end{equation}
and we can obtain the fundamental gap of a system using only one calculation. But from Eqs.\ (\ref{eqIP}), (\ref{eqEA}), (\ref{delxc}), (\ref{eq:FBG}), and (\ref{eq:HOMO-LUMO}), we know that there exists a difference between the fundamental gap and HOMO-LUMO gap,
\begin{equation}
E_{g} = {\Delta}_{KS} + {\Delta}_{xc}.
\end{equation}
As previously mentioned, ${\Delta}_{xc}$ has been shown to be close to zero for LC hybrid functionals \cite{Tsuneda10}, so the HOMO-LUMO gap calculated by a LC hybrid functional should be close to the fundamental gap.

To evaluate the performance of the functionals on fundamental gap, we construct another database called FG115, which shares the same molecular geometries with the EA115 database. For consistency, the reference values 
of fundamental gaps are also obtained via the CCSD(T) calculations described in the last subsection (using Eqs.\ (\ref{defIP}), (\ref{defEA}), and (\ref{eq:FBG})). 

To examine the performance of density functionals, we evaluate the fundamental gaps using three different estimates, with 6-311++G(3df,3pd) basis and EML(75,302) grid. The results are shown from Table \ref{LUHO} to Table \ref{IPEA}, in order of increasing the number of SCF calculations required for each molecule. In the estimate requiring three calculations, the results are similar for the three functionals. $\omega $B97X-D gives worse results than other functionals in the estimate requiring two calculations. In the simplest estimate, the HOMO-LUMO gap, which requires only one SCF calculation for each system, $\omega $M05-D significantly outperforms the other two functionals. The reference values of FG115 and detailed HOMO-LUMO gap results by DFT methods are given in the supplementary material \cite{SI}.

\begin{table}
\caption{\label{LUHO}Statistic errors (in eV) of HOMO-LUMO gaps for the FG115 database. The energy gap of each system is evaluated by only one SCF calculation.}
\begin{ruledtabular}
\begin{tabular}{ccccc}
System & Error & \textit{$\omega $}M05-D & M05-2X & \textit{$\omega $}B97X-D \\ \hline 
atoms  & MSE & -1.14 & -2.56 & -1.55 \\ 
(18) & MAE & 1.43 & 2.56 & 1.79 \\ 
 & rms & 1.62 & 2.79 & 2.05 \\ \hline 
molecules & MSE & -0.62 & -2.00 & -1.15 \\ 
(97) & MAE & 0.73 & 2.00 & 1.15 \\ 
 & rms & 0.93 & 2.13 & 1.34 \\ \hline 
total & MSE & -0.70 & -2.08 & -1.21 \\ 
(115) & MAE & 0.84 & 2.08 & 1.25 \\ 
 & rms & 1.07 & 2.24 & 1.48 \\ 
\end{tabular}
\end{ruledtabular}
\end{table}

\begin{table}
\caption{\label{HOHO}Statistic errors (in eV) of fundamental gaps for the FG115 database, each evaluated by the difference of HOMO energies between the neutral molecule and anion. The energy gap of each system is evaluated by two SCF calculations.}
\begin{ruledtabular}
\begin{tabular}{ccccc}
System & Error & \textit{$\omega $}M05-D & M05-2X & \textit{$\omega $}B97X-D \\ \hline 
atoms& MSE & -0.95 & -0.83 & -1.04 \\ 
(18) & MAE & 0.98 & 0.87 & 1.08 \\ 
 & rms & 1.17 & 1.00 & 1.30 \\ \hline 
molecules & MSE & -0.31 & -0.42 & -0.55 \\ 
(97) & MAE & 0.56 & 0.51 & 0.72 \\ 
 & rms & 0.70 & 0.60 & 0.85 \\ \hline 
total & MSE & -0.41 & -0.48 & -0.63 \\ 
(115) & MAE & 0.62 & 0.57 & 0.78 \\ 
 & rms & 0.79 & 0.68 & 0.93 \\ 
\end{tabular}
\end{ruledtabular}
\end{table}

\begin{table}
\caption{\label{IPEA}Statistic errors (in eV) of IP-EA values for the FG115 database. The energy gap of each system is evaluated by three SCF calculations.}
\begin{ruledtabular}
\begin{tabular}{ccccc}
System & Error & \textit{$\omega $}M05-D & M05-2X & \textit{$\omega $}B97X-D \\ \hline 
atoms & MSE & 0.28 & 0.33 & 0.28 \\ 
(18) & MAE & 0.35 & 0.36 & 0.36 \\ 
 & rms & 0.60 & 0.63 & 0.59 \\ \hline 
molecules & MSE & 0.34 & 0.42 & 0.22 \\ 
(97) & MAE & 0.44 & 0.50 & 0.39 \\ 
 & rms & 0.73 & 0.78 & 0.68 \\ \hline 
total & MSE & 0.33 & 0.40 & 0.23 \\ 
(115) & MAE & 0.43 & 0.48 & 0.39 \\ 
 & rms & 0.71 & 0.75 & 0.66 \\ 
\end{tabular}
\end{ruledtabular}
\end{table}

\subsection{Excitation Energies}

To assess the performance of density functionals on excitation energies, we perform TDDFT calculations on five small molecules \cite{Hirata99}, which include nitrogen gas (\ce{N2}), carbon monoxide (CO), 
water (\ce{H2O}), ethylene (\ce{C2H4}) and formaldehyde (\ce{CH2O}), with 6-311(2+,2+)G** basis and EML(99,590) grid. The molecular geometries, experimental values of excitation energy are taken from Ref.\ \cite{Hirata99}. 
The detail results and mean absolute errors of all excited states are listed in Table \ref{VR}. The new $\omega$M05-D functional yields excellent performance, especially for the Rydberg excitations. Note that $\omega$M05-D 
outperforms $\omega$B97X-D in both HOMO energies and Rydberg excitations, due to the larger fraction of short-range HF exchange included in $\omega$M05-D (both functionals possess the same amount of LR-HF exchange).  

\begin{table*}
\caption{\label{VR}Vertical excitation energies (in eV) of several low-lying excited states of N${}_{2}$, CO, water, formaldehyde and ethylene using 6-311(2+,2+)G** basis set. The geometries and experimental values are taken 
from Ref. \cite{Hirata99}.}
\begin{scriptsize}
\begin{ruledtabular}
\begin{tabular*}{\textwidth}{llrrrr}
Mol. & State & Exp. & \textit{$\omega $}M05-D & M05-2X & \textit{$\omega $}B97X-D \\ \hline
 & V${}^{1}$$\Pi $$_{g}$ & 9.31 & 9.30 & 9.42 & 9.38 \\
 & V${}^{1}$$\Sigma $$^{-}_{u}$ & 9.97 & 8.76 & 8.35 & 9.31 \\ 
 & V${}^{1}$$\Delta $$_{u}$ & 10.27 & 10.14 & 10.51 & 9.82 \\
N${}_{2}$ & V${}^{3}$$\Sigma $$^{+}_{u}$ & 7.75 & 7.86 & 8.30 & 7.17 \\
 & V${}^{3}$$\Pi $$_{g}$ & 8.04 & 7.94 & 8.12 & 7.82 \\
 & V${}^{3}$$\Delta $$_{u}$ & 8.88 & 8.74 & 8.35 & 8.23 \\
 & V${}^{3}$$\Sigma $$^{-}_{u}$ & 9.67 & 8.76 & 9.26 & 9.31 \\
 & V${}^{3}$$\Pi $$_{u}$ & 11.19 & 11.30 & 11.72 & 10.98 \\ \hline
 & V${}^{1}$$\Pi $ & 8.51 & 8.51 & 8.74 & 8.47 \\
 & V${}^{1}$$\Sigma $$^{-}$ & 9.88 & 9.36 & 9.11 & 9.78 \\ 
CO & V${}^{3}$$\Pi $ & 6.32 & 6.66 & 7.03 & 6.07 \\ 
 & V${}^{3}$$\Sigma $$^{+}$ & 8.51 & 8.47 & 8.87 & 8.00 \\
 & V${}^{3}$$\Delta $ & 9.36 & 9.19 & 9.11 & 8.88 \\ 
 & V${}^{3}$$\Sigma $$^{-}$ & 9.88 & 9.36 & 9.55 & 9.78 \\ \hline 
 & R$^{1}$B${}_{1}$ & 7.4 & 7.68 & 8.04 & 7.23 \\ 
 & R${}^{1}$A${}_{2}$ & 9.1 & 9.14 & 9.60 & 8.63 \\ 
H${}_{2}$O & R${}^{1}$A${}_{1}$ & 9.7 & 9.73 & 10.29 & 9.20 \\ 
 & R${}^{1}$B${}_{1}$ & 10.0 & 9.72 & 10.32 & 9.17 \\
 & R${}^{1}$A${}_{1}$ & 10.17 & 10.06 & 10.71 & 9.49 \\ 
 & R${}^{3}$B${}_{1}$ & 7.2 & 7.27 & 7.66 & 6.89 \\ \hline 
 & R${}^{1}$B${}_{3u}$ & 7.11 & 7.53 & 7.61 & 7.02 \\
 & V${}^{1}$B${}_{1u}$ & 7.60 & 7.80 & 8.07 & 7.52 \\
 & R${}^{1}$B${}_{1g}$ & 7.80 & 7.87 & 8.07 & 7.59 \\
 & R${}^{1}$B${}_{2g}$ & 8.01 & 8.15 & 8.19 & 7.66 \\ 
 & R${}^{1}$A${}_{g}$ & 8.29 & 8.36 & 8.52 & 7.87 \\ 
C${}_{2}$H${}_{4}$ & R${}^{1}$B${}_{3u}$ & 8.62 & 8.76 & 8.80 & 8.36 \\
 & V${}^{3}$B${}_{1u}$ & 4.36 & 4.64 & 4.99 & 4.12 \\ 
 & R${}^{3}$B${}_{3u}$ & 6.98 & 7.43 & 7.48 & 6.92 \\
 & R${}^{3}$B${}_{1g}$ & 7.79 & 7.47 & 7.82 & 7.50 \\ 
 & R${}^{3}$B${}_{2g}$ & 7.79 & 8.06 & 8.07 & 7.56 \\  
 & R${}^{3}$A${}_{g}$ & 8.15 & 8.12 & 8.11 & 7.63 \\ \hline 
 & V${}^{1}$A${}_{2}$ & 4.07 & 3.63 & 3.68 & 3.88 \\ 
 & R${}^{1}$B${}_{2}$ & 7.11 & 7.48 & 7.92 & 6.96 \\
 & R${}^{1}$B${}_{2}$ & 7.97 & 8.13 & 8.58 & 7.66 \\
 & R${}^{1}$A${}_{1}$ & 8.14 & 9.13 & 9.47 & 8.74 \\
 & R${}^{1}$A${}_{2}$ & 8.37 & 8.30 & 8.84 & 7.84 \\ 
CH${}_{2}$O & R${}^{1}$B${}_{2}$ & 8.88 & 8.86 & 9.21 & 8.52 \\
 & V${}^{3}$A${}_{2}$ & 3.50 & 3.02 & 3.12 & 3.21 \\ 
 & V${}^{3}$A${}_{1}$ & 5.86 & 5.70 & 6.02 & 5.29 \\ 
 & R${}^{3}$B${}_{2}$ & 6.83 & 7.33 & 7.74 & 6.81 \\ 
 & R${}^{3}$B${}_{2}$ & 7.79 & 7.97 & 8.37 & 7.50 \\
 & R${}^{3}$A${}_{1}$ & 7.96 & 8.06 & 8.45 & 7.56 \\  \hline 
MAE & Valence &  & 0.31 & 0.46 & 0.32 \\ 
 & Rydberg &  & 0.22 & 0.47 & 0.35 \\
\end{tabular*}
\end{ruledtabular}
\end{scriptsize}
\end{table*}

\begin{figure}
\includegraphics[width=15cm]{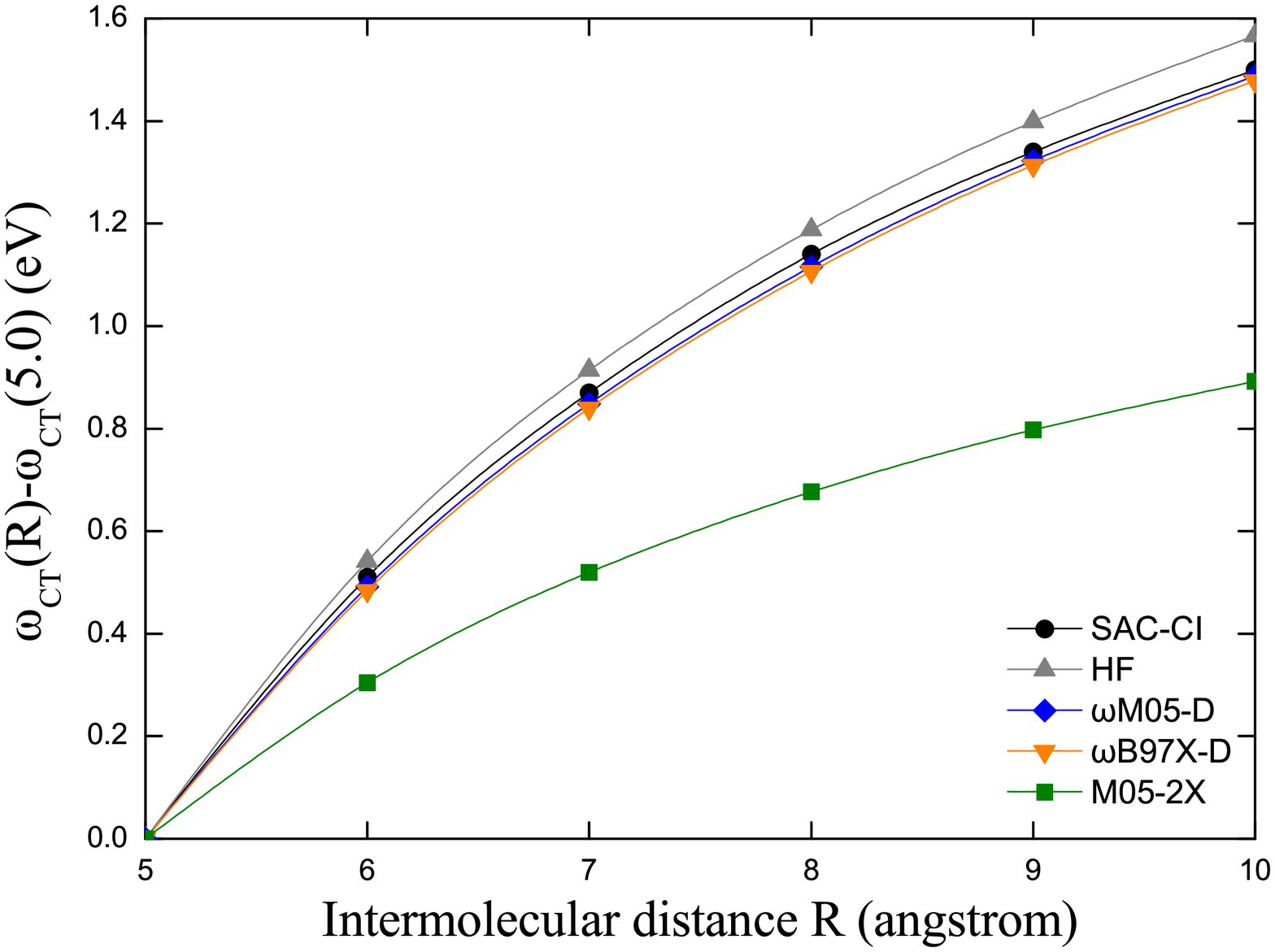}
\caption{The lowest CT excitation of \ce{C2H4}$\cdots$\ce{C2F4} dimer along the intermolecular distance $R$ (in $\text{\AA}$). For all functionals, the excitation at 5 $\text{\AA}$ is set to zero.}
\label{CT1}
\end{figure}

\begin{figure}
\includegraphics[width=15cm]{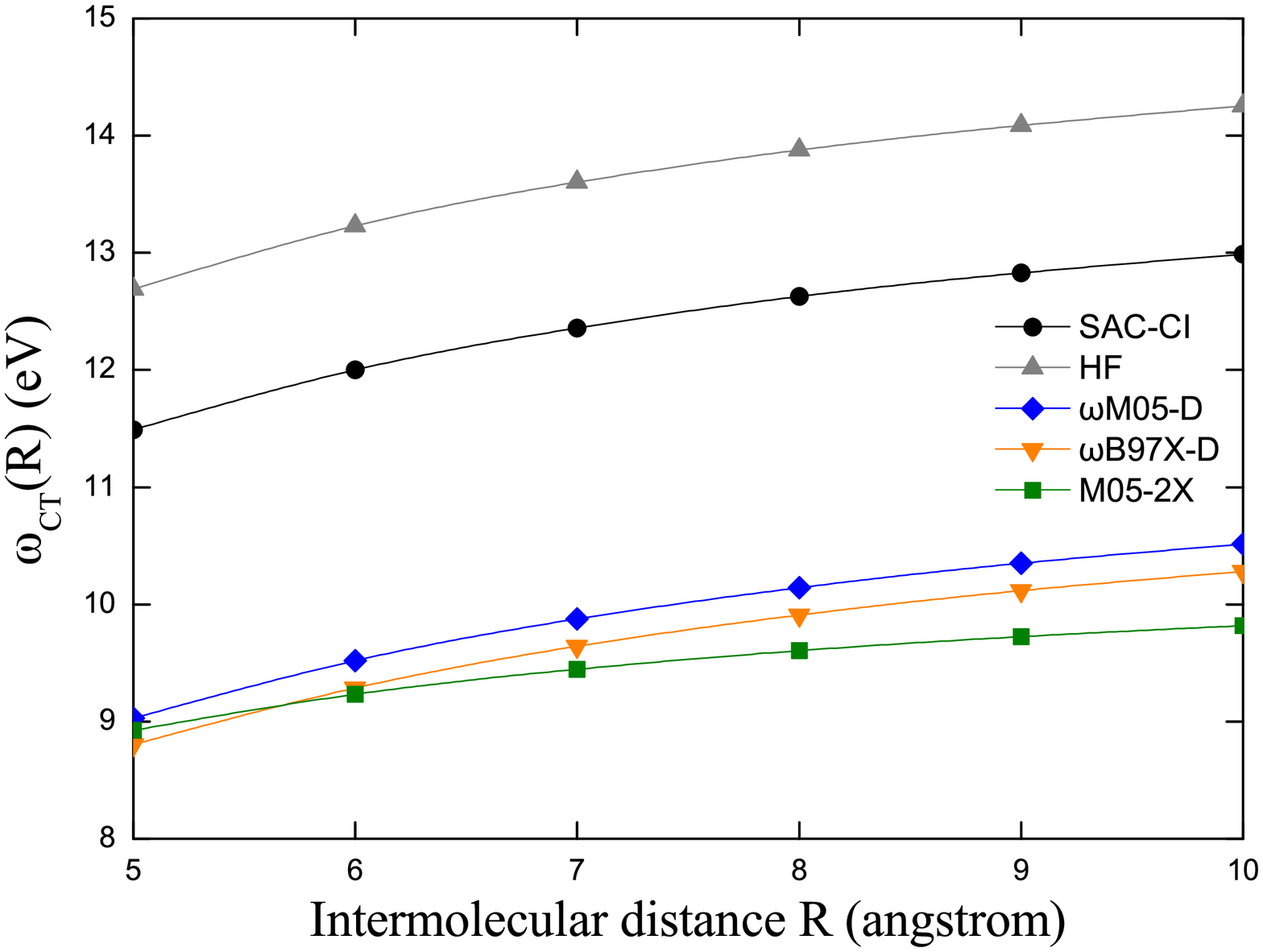}
\caption{The lowest CT excitation of \ce{C2H4}$\cdots$\ce{C2F4} dimer along the intermolecular distance $R$ (in $\text{\AA}$).}
\label{CT2}
\end{figure}

Following Dreuw {\it et al.}, we perform TDDFT calculations for the lowest charge-transfer (CT) excitation between ethylene and tetrafluroethylene, with a separation of $R$. Dreuw {\it et al.} have shown that the exact CT 
excitation energy from the HOMO of donor to the LUMO of acceptor should have the following asymptote \cite{Dreuw03}:
\begin{equation}
\omega_\text{CT}(R \rightarrow \infty) \approx -\frac{1}{R}+\text{IP}_\text{D}-\text{EA}_\text{A},
\end{equation}
where $\text{IP}_\text{D}$ is the ionization potential of donor and $\text{EA}_\text{A}$ is the electron affinity of acceptor. Fig.\ \ref{CT1} shows the trend of the excitation curves, and indicates the LC hybrid functionals obviously 
outperforms the global hybrid M05-2X. For the values of the excitation energies, as shown in Fig.\ \ref{CT2}, $\omega$M05-D is about 0.2 electron volt better than $\omega$B97X-D.

\section{CONCLUSIONS}

We have developed a LC hybrid MGGA-D functional, called $\omega$M05-D, which includes 100\% long-range exact exchange, a fraction ($\approx$ 37\%) of short-range exact exchange, a modified M05 
exchange density functional for short-range interaction, the M05 correlation density functional \cite{Zhao05,Zhao06}, and empirical atomic-pairwise dispersion corrections. 
For the modified short-range M05 exchange density functional, we have investigated two models. After comparisons in the training set and test sets, we decide to propose the one based on our new 
LC scheme $\omega$M05-D, and marked the trial one as $\omega$M05s-D. When the constraint of $\omega$ = 0 is applied, $\omega$M05-D and $\omega$M05s-D are both reduced to the existing M05 functional form \cite{Zhao05,Zhao06} with the same empirical atomic-pairwise dispersion corrections. The constrained form ($\omega$ = 0), when re-optimized on the same training set, provides worse performance on the training set, indicating 
that the single extra degree of freedom corresponding to long-range exchange is of physical significance to a hybrid MGGA. 

Since $\omega$M05-D is a parametrized functional, we test it against the trial simple model $\omega$M05s-D as well as two closely related functionals (M05-2X \cite{Zhao06} and $\omega$B97X-D \cite{Chai08_2}) on a 
separate independent test set of data, which includes further atomization energies, reaction energies, noncovalent interaction energies, equilibrium geometries, energy curve for homonuclear diatomic cation dissociations, 
frontier orbital energies and fundamental gaps. The three databases assessing frontier orbital energies and fundamental gaps are presented for the first time. For excitation energies, we calculate valence and Rydberg 
excitations, as well as a charge-transfer excited state. Compared to $\omega$M05s-D, noticeable difference in transferability for atomization energies largely decides our proposed model. $\omega$M05-D consistently 
outperforms M05-2X (and performs comparably to $\omega$B97X-D) on the test sets, and shows smaller SIE and better asymptotic behavior relative to both the M05-2X and $\omega$B97X-D.

\begin{acknowledgments}
This work was supported by National Science Council of Taiwan (Grant No. NSC98-2112-M-002-023-MY3), National Taiwan University (Grant No. 99R70304 and 10R80914-1), and NCTS of Taiwan. 
\end{acknowledgments}

\newpage

\section*{Supplementary material}

\begin{longtable*}{lrrrr}
\caption{\label{table:IP132_HOMO} -$\epsilon_N(N)$ (in eV) of IP131 database. Most experimental reference values are collected from the NIST database, \cite{NIST} while other publications \cite{Gill92,Brundle72,Niessen82,Colbourne78,Frost72,Bieri82,Cvitas77,Bieri81,Asbrink80,Gelius71,Asbrink81,Bieri80,Niessen80} are adopted for the molecules marked.}\\
\hline \hline
	Molecule	&	Reference	&	~~~$\omega$M05-D	&	~~~M05-2X	&	~~~$\omega$B97X-D	\\ \hline
	H (Hydrogen atom)	&	13.60	&	11.10	&	10.44	&	11.04	\\
	He (Helium atom)	&	24.59	&	20.75	&	20.75	&	20.18	\\
	Li (Lithium atom)	&	5.39	&	5.06	&	4.19	&	5.22	\\
	Be (Beryllium atom) \cite{Gill92}	&	9.32	&	8.19	&	7.29	&	8.27	\\
	B (Boron atom)	&	8.30	&	7.21	&	6.39	&	7.24	\\
	C (Carbon atom)	&	11.26	&	9.74	&	9.11	&	9.57	\\
\hline \hline
\end{longtable*}

\begin{longtable*}[l]{lrrrrrrr}
\caption{\label{table:EA115_HOMO} -$\epsilon_{N+1}(N+1)$ and -$\epsilon_{N+1}(N)$ (in eV) of EA115 database. Reference values are calculated by the CCSD(T) theory.}\\
\hline \hline
&	reference	&	\multicolumn{2}{c}{$\omega$M05-D}			&	\multicolumn{2}{c}{M05-2X}			&	\multicolumn{2}{c}{$\omega$B97X-D}			\\	
		molecule	&		&	-$\epsilon_{N+1}(N+1)$	&	-$\epsilon_{N+1}(N)$	&	-$\epsilon_{N+1}(N+1)$	&	-$\epsilon_{N+1}(N)$	&	-$\epsilon_{N+1}(N+1)$	&	-$\epsilon_{N+1}(N)$	\\	\hline
	H (Hydrogen atom)		&	0.75	&	0.84	&	-0.52 	&	-0.08	&	-0.53 	&	0.84	&	-0.53 	\\	
	He (Helium atom)		&	-2.63	&	-4.35	&	-4.76 	&	-5.17	&	-3.89 	&	-4.27	&	-4.30 	\\	
	Li (Lithium atom)		&	0.62	&	0.68	&	-0.44 	&	0.00	&	0.14 	&	0.71	&	-0.19 	\\	
	Be (Beryllium atom)		&	-0.36	&	-0.46	&	-0.52 	&	-1.20	&	0.35 	&	-0.41	&	-0.30 	\\	
	B (Boron atom)		&	0.25	&	0.19	&	0.38 	&	-0.65	&	1.28 	&	0.16	&	0.57 	\\
\hline \hline
\end{longtable*}

\begin{longtable*}{lrrrr}
\caption{\label{table:G53_HOMOLUMOgap} Reference values of FG115 database calculated by the CCSD(T) theory, compared with HOMO-LUMO gap results by DFT methods.}\\
\hline \hline
	Molecule		&	Reference	&	~~~$\omega$M05-D	&	~~~M05-2X	&	~~~$\omega$B97X-D	\\ \hline
	H (Hydrogen atom)	&	12.86	&	11.62	&	10.98	&	11.57	\\
	He (Helium atom)	&	27.23	&	25.57	&	24.70	&	24.56	\\
	Li (Lithium atom)	&	4.22	&	5.49	&	4.05	&	5.41	\\
	Be (Beryllium atom)	&	9.66	&	8.70	&	6.94	&	8.57	\\
	B (Boron atom)	&	7.99	&	6.83	&	5.11	&	6.66	\\
	C (Carbon atom)	&	9.97	&	8.05	&	6.69	&	7.59	\\
\hline \hline
\end{longtable*}
\pagebreak

\begin{longtable*}{p{0.25\textwidth}m{0.25\textwidth}m{0.25\textwidth}m{0.25\textwidth}}
\caption{\label{table:IP133} Experimental molecular geometries (in angstrom) for IP131, EA115 and FG115 databases collected from the NIST database \cite{NIST}. The  integers following the molecule are in order the spin multiplicities of neutral species, cation and anion. The molecules with only the spin multiplicity of the neutral species shown are in neither EA115 nor FG115 database. The spin multiplicities of neutral species are experimental values specified in the NIST database \cite{NIST}. Each spin multiplicity of cations and anions is decided by CCSD(T)/aug-cc-pVTZ calculations, and corresponds to the lowest energy in between various spin multiplicities.}\\

\hline
\hline

\multicolumn{4}{l}{$\ce{H}$ (Hydrogen atom)~~~2~~~Null~~~1}\\
H\\
\hline

\multicolumn{4}{l}{$\ce{He}$ (Helium atom)~~~1~~~2~~~2}\\
He\\
\hline

\multicolumn{4}{l}{$\ce{Li}$ (Lithium atom)~~~2~~~1~~~1}\\
Li\\
\hline

\multicolumn{4}{l}{$\ce{Be}$ (Beryllium atom)~~~1~~~2~~~2}\\
\hline
\hline
\end{longtable*}
\pagebreak

\begin{table}
\caption{Atomic energies (in kcal/mol) from the H atom to the Ar atom \cite{Chakravorty93}.}
\begin{ruledtabular}
\begin{tabular}{crrrrrr}
Atom & Reference & \textit{$\omega $}M05-D &  M05-2X & \textit{$\omega $}B97X-D \\ \hline
H & -313.76  & -311.52  &   -313.31  & -315.51  \\ 
He & -1822.12  & -1822.11  &   -1823.25  & -1824.57  \\ 
Li & -4692.56  & -4693.72  &   -4694.37  & -4699.80  \\ 
Be & -9203.92  & -9198.97  &   -9204.50  & -9202.60  \\ 
B & -15470.58  & -15466.17  &   -15471.98  & -15468.00  \\ 
C & -23748.12  & -23746.17  &   -23752.06  & -23745.12  \\ 
N & -34255.28  & -34254.32  &   -34260.32  & -34251.37  \\ 
O & -47105.50  & -47107.44  &   -47113.72  & -47103.64  \\ 
F & -62584.04  & -62587.59  &   -62593.80  & -62583.06  \\ 
Ne & -80909.66  & -80910.31  &   -80917.57  & -80906.81  \\ 
Na & -101816.42  & -101814.09  &   -101820.46  & -101818.06  \\ 
Mg & -125535.30  & -125538.45  &   -125549.12  & -125538.97  \\ 
Al & -152074.59  & -152077.17  &   -152084.44  & -152078.77  \\ 
Si & -181575.73  & -181577.53  &   -181581.20  & -181577.86  \\ 
P & -214143.51  & -214141.04  &   -214140.76  & -214140.27  \\ 
S & -249818.09  & -249817.74  &   -249815.58  & -249817.21  \\ 
Cl & -288747.57  & -288747.04  &   -288743.15  & -288746.55  \\ 
Ar & -331036.74  & -331035.35  &   -331028.07  & -331036.25  \\ \hline
MSE &  & 0.37  &   -3.01  & -0.05  \\ 
MAE &  & 2.02  &   5.10  & 2.51  \\ 
rms &  & 2.42  &   6.31  & 2.96  \\ 
\end{tabular}
\end{ruledtabular}
\end{table}
\clearpage

\begin{table*}
\caption{Non-hydrogen transfer barrier heights (in kcal/mol) of the NHTBH38/04 set \cite{Zhao_JPCA05,*Zhao_E06}.}
\begin{scriptsize}
\begin{ruledtabular}
\begin{tabular*}{\textwidth}{lcrrrrrr}
Reactions &  & $\Delta $\textit{E${}_{ref}$} & \textit{$\omega $}M05-D &   M05-2X & \textit{$\omega $}B97X-D \\ \hline
\textbf{Heavy-atom transfer reactions} &  &  &  &  &      \\ 
\ce{H + N2O $\rightarrow$ OH + N2} & \textit{V${}^{f}$} & 18.14 & 16.52  &   18.82  & 17.47  \\ 
 & \textit{V${}^{r}$} & 83.22 & 80.54  &   85.45  & 77.75  \\ 
\ce{H + FH $\rightarrow$ HF + H} & \textit{V${}^{f}$} & 42.18 &38.34  &   41.16  & 40.56  \\ 
 & \textit{V${}^{r}$} & 42.18 & 38.34 &   41.16  & 40.56  \\ 
\ce{H + ClH $\rightarrow$ HCl + H} & \textit{V${}^{f}$} & 18.00 & 17.83 &   19.64  & 18.19  \\ 
 & \textit{V${}^{r}$} & 18.00 & 17.83  &   19.64  & 18.19  \\
\end{tabular*}
\end{ruledtabular}
\end{scriptsize}
\end{table*}
\clearpage

\begin{table*}
\caption{Hydrogen transfer barrier heights (in kcal/mol) of the HTBH38/04 set \cite{Zhao04,Zhao_JPCA05,*Zhao_E06}.}
\begin{scriptsize}
\begin{ruledtabular}
\begin{tabular*}{\textwidth}{lcrrrrrr}
Reactions &  & $\Delta E{}_{ref}$ & $\omega $M05-D &   M05-2X & $\omega $B97X-D \\ \hline
\ce{H + HCl $\rightarrow$ H2 + Cl} & ${V}^{f}$ & 5.7 & 3.61  &   4.82  & 4.08  \\ 
 & ${V}^{r}$ & 8.7 & 4.68  &   5.94  & 4.58  \\ 
\ce{OH + H2 $\rightarrow$ H + H2O} & ${V}^{f}$ & 5.1 & 1.94  &   4.61  & 2.22  \\ 
 & ${V}^{r}$ & 21.2 & 18.69  &   21.90  & 18.77  \\ 
\ce{CH3  + H2  $\rightarrow$ H + CH4} & ${V}^{f}$ & 12.1 & 9.14  &   11.72  & 9.06  \\ 
 & ${V}^{r}$ & 15.3 & 13.35  &   16.07  & 13.54  \\ 
\ce{OH + CH4 $\rightarrow$ CH3 + H2O} & ${V}^{f}$ & 6.7 & 2.53  &   4.69  & 3.04  \\
\end{tabular*}
\end{ruledtabular}
\end{scriptsize}
\end{table*}
\clearpage

\begin{table*}
\caption{Interaction energies (in kcal/mol) for the S22 set with new reference values. \cite{Marshall11} The counterpoise corrections are used to reduce the basis set superposition errors. Monomer deformation energies are not included.}
\begin{scriptsize}
\begin{ruledtabular}
\begin{tabular*}{\textwidth}{lrrrrrr}
Complex [Symmetry] & $\Delta E{}_{ref}$ & $\omega $M05-D &  M05-2X & $\omega $B97X-D \\ \hline
\textbf{Hydrogen bonded complexes} &  &  &  &    \\ 
\ce{(NH3)2  [C$_{2h}$]} & -3.133 & -3.00  &  -3.12  & -3.08  \\ 
\ce{(H2O)2  [C$_s$]} & -4.989 & -4.89  &  -5.08  & -4.97  \\ 
Formic acid dimer [C${}_{2h}$] & -18.753 & -18.98  &  -19.41  & -19.28  \\ 
Formamide dimer [C${}_{2h}$] & -16.062 & -15.65  &  -15.81  & -16.17  \\ 
Uracil dimer [C${}_{2h}$] & -20.641 & -19.90  &  -19.64  & -20.43  \\ 
2-pyridoxine$\cdot$2-aminopyridine [C$_1$] & -16.934 & -16.40  &  -15.36  & -17.06  \\
\end{tabular*}
\end{ruledtabular}
\end{scriptsize}
\end{table*}
\clearpage

\begin{table*}
\caption{Comparison of errors of different functionals for the reaction energies (in kcal/mol) of the 30 chemical reactions in the NHTBH38/04 and HTBH38/04 database \cite{Zhao04,Zhao_JPCA05,*Zhao_E06}.}
\begin{scriptsize}
\begin{ruledtabular}
\begin{tabular*}{\textwidth}{lrrrrrr}
Reactions & $\Delta E{}_{ref}$ & $\omega $M05-D & M05-2X & $\omega $B97X-D \\ \hline
\ce{H + N2O $\rightarrow$ OH + N2} & -65.08 & -64.03  &  -66.63  & -60.28  \\ 
\ce{H + FCH3$\rightarrow$ HF + CH3} & -26.64 & -25.50  &  -23.58  & -24.44  \\ 
\ce{H + F2 $\rightarrow$ HF + F} & -103.91 & -106.58  &  -111.74  & -104.56  \\ 
\ce{CH3  + FCl$\rightarrow$ CH3F + Cl} & -52.74 & -53.41  &  -55.90  & -53.79  \\ 
F${}^{-}$  + \ce{CH3Cl $\rightarrow$ FCH3}  + Cl${}^{-}$ & -32.65 & -36.03  &  -37.75  & -34.84  \\ 
F${}^{-}$$\cdots$\ce{CH3Cl $\rightarrow$ FCH3}$\cdots$Cl${}^{-}$ & -26.73 & -28.71  &  -29.84  & -27.43  \\ 
OH${}^{-}$ + \ce{CH3F $\rightarrow$ HOCH3} + F${}^{-}$ & -20.11 & -20.02  &  -19.76  & -21.21  \\
\end{tabular*}
\end{ruledtabular}
\end{scriptsize}
\end{table*}

\begin{table*}
\tiny
\caption{Binding energies (in kcal/mol) of several sets of noncovalent interactions. The first three sets are taken from Ref. \cite{Zhao_JPCA05_2} with monomer deformation energies taken into considerations. 
The last three sets are taken from Ref. \cite{Jurecka_PCCP06} without considering monomer deformation energies. The counter-point corrections are applied for all the cases.}
\begin{ruledtabular}
\begin{tabular*}{\textwidth}{lrrrrrr}
Complex & $\Delta $\textit{E${}_{ref}$} & \textit{$\omega $}M05-D &  M05-2X & \textit{$\omega $}B97X-D \\ \hline
\textbf{Charge-transfer complexes} &  &  &  &  \\ 
\ce{C2H4$\cdots$F2} & 1.06 & 0.79 &  1.00 & 0.78 \\ 
\ce{NH3$\cdots$F2} & 1.81 & 1.41 &  1.44 & 1.50 \\ 
\ce{C2H2}$\cdots$ClF & 3.81 & 3.82 &  4.51 & 3.66 \\ 
HCN$\cdots$ClF & 4.86 & 4.21 &  4.72 & 4.21 \\ 
\ce{NH3$\cdots$Cl2} & 4.88 & 4.61 &  5.02 & 4.81 \\ 
\ce{H2O}$\cdots$ClF & 5.36 & 5.21 &  5.83 & 5.18 \\
\end{tabular*}
\end{ruledtabular}
\end{table*}
\clearpage


\end{document}